# Brain tumour genetic network signatures of survival

James K. Ruffle FRCR MSc[1], Samia Mohinta MSc[1], Guilherme Pombo MSc[1], Robert Gray PhD[1], Valeriya Kopanitsa MBBS BSc[1], Faith Lee BSc[1], Sebastian Brandner MD FRCPath[2], Harpreet Hyare FRCR PhD[1], and Parashkev Nachev FRCP PhD[1]

[1]*Queen Square Institute of Neurology, University College London, London WC1N 3BG, UK*

[2]*Division of Neuropathology and Department of Neurodegenerative Disease, Queen Square Institute of Neurology, University College London, London WC1N 3BG, UK*

Running title:

Brain tumour genetic network signatures of survival

Correspondence to:

Dr James K Ruffle

Email: j.ruffle@ucl.ac.uk

Address: Institute of Neurology, UCL, London WC1N 3BG, UK

Correspondence may also be addressed to:

Professor Parashkev Nachev

Email: p.nachev@ucl.ac.uk

Address: Institute of Neurology, UCL, London WC1N 3BG, UK

Abbreviations: Amp, amplification; GBM, glioblastoma multiforme; Hist, histone; MCMC, Markov Chain Monte Carlo; Methyl, methylation; Mut, mutant; PCA, principal component analysis; Ret, retained; WAIC, widely applicable information criterion; WHO, World Health Organisation; WT, wild-type.

Funding: JKR was supported by the Guarantors of Brain, the Medical Research Council (MR/X00046X/1), the NHS Topol Digital Fellowship and the UCL CDT i4health. PN is supported by the Wellcome Trust (213038/Z/18/Z). PN, SB, and HH are supported by the UCLH NIHR Biomedical Research Centre.

Conflict of interest: The authors declare that there are no competing interests.

Authorship: Conceptualization: JR, HH, SB, PN; Data: JR, VK, FL, HH, SB, PN; Methodology: JR, PN; Software: JR, SM, GP, RG; Validation: JR, PN; Formal analysis: JR, SM, GP; Manuscript writing JR, PN; Manuscript reviewing, and editing: JR, SM, VK, FL, GP, RG, HH, SB, PN. All authors have been involved in the writing of the manuscript and have read and approved the final version.

Data sharing: All code shall be made publicly available upon publication at https://github.com/high-dimensional. Trained model weights are available upon request. Patient data is not available for dissemination under the ethical framework that governs its use.

# Abstract

Tumour heterogeneity is increasingly recognized as a major obstacle to therapeutic success across neuro-oncology. Gliomas are characterised by distinct combinations of genetic and epigenetic alterations, resulting in complex interactions across multiple molecular pathways. Predicting disease evolution and prescribing individually optimal treatment requires statistical models complex enough to capture the intricate (epi)genetic structure underpinning oncogenesis. Here, we formalize this task as the inference of distinct patterns of connectivity within hierarchical latent representations of genetic networks. Evaluating multi-institutional clinical, genetic, and outcome data from 4023 glioma patients over 14 years, across 12 countries, we employ Bayesian generative stochastic block modelling to reveal a hierarchical network structure of tumour genetics spanning molecularly confirmed glioblastoma, IDH-wildtype; oligodendroglioma, IDH-mutant and 1p/19q codeleted; and astrocytoma, IDH-mutant. Our findings illuminate the complex dependence between features across the genetic landscape of brain tumours, and show that generative network models reveal distinct signatures of survival with better prognostic fidelity than current gold standard diagnostic categories.

# Introduction

Brain tumours remain remarkably resistant to treatment, and impose a socioeconomic burden second amongst cancers only to breast and lung[1]. Fewer than half of people with the commonest malignant type—glioblastoma, IDH-wildtype—survive a year, a prognosis unchanged over the past three decades in the face of an increase in incidence by more than a sixth[2,3]. These striking numbers suggest fundamental obstacles to treatment success that may signal the need for a radical change in our approach.

One of the greatest obstacles for innovation across oncology is inter- and intra-tumour heterogeneity[4-7]: the presence of richly structured diversity, either between different tumours or within different parts of the same one. Brain tumours typically exhibit numerous genetic mutations, spanning several cellular pathways, that open *multiple* avenues to oncogenesis no *single* intervention could conceivably block. It is no surprise that patients with higher levels of tumour heterogeneity—ranging across genetic[4,8,9], epigenetic, cellular and imaging characteristics[10,11]—exhibit both poorer clinical outcomes and weaker responses to therapy[5,12-14].

A pre-requisite to overcoming heterogeneity is obtaining a structured description, as comprehensive as available data allow, of the parameter space that encloses it. Such a description is difficult to derive because tumour heterogeneity is distributed across many potentially interacting features[4-6], inhabiting a large, high-dimensional parameter space. It requires highly expressive mathematical models, capable of capturing multiple, richly interacting factors. The task is formally representation learning: deriving data-driven, succinct, *signature* representations of tumour genetic features that illuminate their complex inter-relations.

Graph theory provides a powerful approach to such modelling, combining expressivity with intuitive intelligibility[15-22]. It treats the characteristics of a system under study as the nodes of a network, and their relations as the connections, or edges, between nodes. In the context of brain tumours, the nodes correspond to clinical, radiological, histological, or genetic features—applied to distinct tumour properties or to the patient as a whole—and the edges correspond to their pairwise relations. Represented as a graph, the heterogeneous structure of tumours is expressed as distinct regional patterns of connectivity, defining blocks or communities of similarly connected nodes. Organised by a domain of interest—tumour genetics, for example—the inferred communities reveal heterogeneous inter-relations that may underpin oncogenesis and suggest avenues for treatment innnovation[5,18,23,24]. Organised by individual patients, the inferred communities identify patient subpopulations with similar oncological signatures that may signal decisive differences in disease mechanisms, evolution, or treatment response. The novelty of the approach resides in the ability to model complex *interactions* between features that may illuminate mechanistic and prognostic relations opaque to models of the same features taken alone or only in linear combination. Insights may thus be gleaned from routinely collected variables that are familiar individually but unexplored in their collective interactions.

The fidelity of any representational model is constrained by two interacting factors: data sampling—both density and coverage—and the expressivity of the model architecture. Given the manifest complexity of oncogenic mechanisms[23,24], data scale and range will always be limiting, and model architecture will be placed under great stress. Focusing on tumour genetics, we therefore analyse to our knowledge the largest fully-inclusive collection of brain tumour data, spanning 4023 patients, acquired over 14 years, across 12 countries. We exploit recent advances in non-parametric Bayesian generative models of the modular structure of graphs[25,26] to derive robust hierarchical latent genetic representations, yielding a set of signature network patterns that illuminate potentially critical relations between genetic features, and enable finer patient stratification than current diagnostic classification systems allow. We demonstrate, quantitatively, the utility of our representations by comparing their fidelity in predicting survival against both diagnostic labels from the latest World Health Organisation (WHO) brain tumour classification[27] and raw genetic and epigenetic features. Our evaluation shows that a graph approach powered by large-scale, fully-inclusive data can successfully capture tumour genetic heterogeneity, delivering higher fidelity prediction of survival than current representations, and opening the way to richly multimodal generative modelling of the complex landscape of neuro-oncology.

## Materials and methods

### Data

The demographic, procedural, histopathological, tumour genetic, and diagnostic labels of 9518 neuro-oncology patients referred to our national centre were recorded prospectively from 2006 to 2020 (Figure 1, Supplementary Figure 1). The distribution of countries of origin was, in descending order, the UK (n=9149), Colombia (n=170), Sweden (n=157), Latvia (n=14), Hungary (n=6), France (n=4), Lithuania (n=3), USA (n=3), Republic of Ireland (n=2), and Portugal, Bulgaria, Bermuda, Spain, Malta, Greece, South Africa, Poland, India and Peru (all n=1). Of the UK-resident patients, 3134 were managed at our institution, leaving 6384 elsewhere. The data included age, sex, referral date, biopsy/surgical resection date, histology-informed diagnosis in accordance with the current WHO CNS5 classification of brain tumours[27], and the status of tumour genetic features recorded as part of routine clinical care. A total of 7809 patients identified as having received a final diagnosis of glioma. We removed patients with partial molecular panel results and/or lacking WHO CNS5 diagnostic information to yield a cohort of n=4023. The surveyed genetic features included IDH (isocitrate dehydrogenase), ATRX (ATP-dependent helicase ATP-dependent helicase), TERT (telomerase reverse transcriptase), histone, BRAF (proto-oncogene B-Raf) point or fusion, chromosomal 1p/19q deletions, MGMT (O-6-methylguanine-DNA methyltransferase) percentage methylation, and degree (if any) of EGFR (epidermal growth factor receptor) amplification. Extent of EGFR amplification (if any) were stratified into low (1-7 copies), medium (8-15 copies), and high (16 copies or more), in accordance with standard practice at our centre. Sampling frequency, including missing data, was captured within the stochastic block model itself, exploiting its generative nature.

Our focus is on quantifying the potential value of graph-theoretic analysis of data routinely acquired during standard clinical care. Such an approach lowers the barriers to real-world application, for no change to standard investigational pathways is required, and enables the derivation of insights from historical data. Neither additional time nor economic cost is incurred, to patient or healthcare provider: the only necessary resource is compute. We therefore included all genetic features acquired as part of routine neuro-oncological care in the molecular neuropathology panel work-up at our centre. This panel is described online[28], and aligns with established clinical practice providing both classification of tumours within the current WHO classification system[27] and prognostic or prescriptive utility such as MGMT methylation status for Temozolomide use[29]. Since our cohort dates back to 2006, it does not include CDKNDA/B testing first recommended in 2018[30], and even now considered optional by many[27]. Our centre did not routinely test for the supplementary IDH mutant astrocytoma diagnostic marker TP53, favouring ATRX instead as recommended by the recent WHO classification[27]. Histological grade, determined by microscopic rather than molecular features, was not available. We did not perform any prior feature selection, for we are interested in the interactions between features our graph technique is specifically designed to illuminate, instead including all data available from our neuro-oncology service. Our modelling approach can cope with high-dimensional data, so no selection is required on methodological grounds.

Our analysis focussed on four major categories of glioma: glioblastoma, IDH-wildtype; astrocytoma, IDH-mutant; oligodendroglioma, IDH-mutant and 1p/19q codeleted; and a final group titled 'other glioma' that combined rarer entities. The rationale for the latter group was that we were wary of drawing inference from much smaller samples of rarer lesions, contrasted to the remaining diagnoses with significantly greater samples sizes, that would otherwise render the performance inequitable[31] (Table 1). Survival data was available for 1323 patients, constrained only by the mechanism of referral. For statistical modelling, we discarded samples where any graph community, diagnostic or genetic variable received fewer than 5 patients, and clamped days of survival at the 1st and 99th percentile to attenuate the influence of extreme outliers. A full cohort breakdown, including where applicable data missingness, is detailed in Table 1. A study flow chart is provided as Supplementary Figure 1.

### Ethical approval

The study was approved by the local ethics committee at University College London. We received ethical permission for the consentless analysis of irrevocably anonymized data collected during routine clinical care.

### Analytic compliance

All analyses were performed and reported in accordance with international TRIPOD and PROBAST-AI guidelines[32].

## Demographic analysis

One-way analysis of variance (ANOVA) with Tukey's procedure was used to establish the relation between patient age and diagnosis, and multivariate logistic regression for patient sex and diagnosis. Our criterion for statistical significance was a family-wise error rate (FWER) adjusted p<0.05, and all p values reported are corrected accordingly. Model coefficients were converted into odds or hazard ratios where appropriate.

## Network genetic signature analysis

A network representation of tumour genetics can be formulated in two ways: with respect to genetic features, yielding signatures of characteristic patterns of genetic lesion co-occurrence, or with respect to patients, yielding distinct subpopulations exhibiting similar genetic signatures. The former illuminates the mechanisms of oncogenesis, the latter their heterogeneous manifestation across the population.

## Stochastic block modelling of tumour genetic inter-relations

The relations between genetic features may be naturally formulated in terms of Bayes' rule[33,34]:

$$P(A|B) = \frac{P(B|A) \cdot P(A)}{P(B)},$$

where $P(A)$ and $P(B)$ refer to the probabilities of the states of given genetic features $A$ and $B$ respectively. $P(B|A)$ is the conditional probability of $B$ given $A$, and $P(A|B)$ is the posterior conditional probability of $A$ given $B$. In general $P(A|B) \neq P(B|A)$ so, unlike merely correlative indices, conditional probabilities enable us to construct a directed probabilistic graph of the pairwise relations between genetic features. The number of edges is given as the number of nodes choose 2, multiplied by two to cover bidirectional conditional probabilities, $2\binom{N}{2}$. We reviewed the weighted edge histogram of the graph according to conditional probability $P(A|B)$, comparing it to arguably simpler metric approximating covariance, the probability of intersection $P(A \cap B)$. Conditional probability edges showed far greater weight variance (range 0.00 to 1.00, ± standard deviation (SD) 0.25) compared to intersection weights (range 0.00 to 0.51, ± SD 0.06). A reasonable assumption drawn from this process were that the use of directed conditional probability weights between genetic features may offer more sophisticated variation of information than simpler intersection (or covariance-based) metrics, and thus were adopted for subsequent mathematical modelling between genes. In compliment, a patient linkage graph was modelled with multi-variately weighted edges by binomial linkage of individual tumour genetic characteristics (schematic for both approaches shown in Figure 2). We characterised simple centrality measures – eigenvector, hub, authority, betweenness, and page rank – weighted by the conditional probability assigned to the directed edges. We then statistically compared these centrality metrics between genetic domains with one-way analysis of variance (ANOVA).

A stochastic block model is a generative model of the community structure of a graph composed of $N$ nodes, divided into $B$ blocks with edges $e_{rs}$ between blocks $r$ and $s$[35]. The model can be framed hierarchically, where edge counts $e_{rs}$ form block multigraphs with nodes corresponding to individual blocks and edge counts arising as edge multiplicities between block pairs, including self-loops. We seek to infer the most plausible partition $\{b_i\}$ of the nodes, where $\{b_i\} \in [1, B]^N$ identifies the block membership of node $i$ in observed network $G$, with maximisation of the posterior likelihood $P(G|\{b_i\})$. The result is a hierarchically organised community structure of nodes assigned into blocks that yields the most compact representation of the graph, as indexed by its minimum description length[36], $\Sigma$. The general approach is described in further detail elsewhere[35].

We can use a layered formulation[26] of the model to distinguish between two potentially conflicting effects: associations between features driven by clinically-directed sampling vs by biologically-driven conditional probability. Key here is formal comparison between models that encode these effects separately, within their own layers, vs those where the distinction is not respected. In a Bayesian setting[26], the procedure for model selection amounts to finding the model parameters, $\{\theta\}$, that maximise the posterior likelihood as

$$P(\{\theta\}|\{G_l\}) = \frac{P(\{G_l\}|\{\theta\})P(\{\theta\})}{P(\{G_l\})},$$

In our case, $\{\theta\} = \{\{b_i\}, \{e_{rs}^l\}\}$, where $N$ nodes are divided into $B$ blocks via the membership vector $\{b_i\} \in [1, B]^N$, and the distribution of covariates in edges in groups $r$ and $s$ is given by the edge counts $e_{rs}$, with $e_{rs}^l$ corresponding to the former at a given layer. $P(\{\theta\})$ is the prior probability of these parameters, with $P(\{G_l\})$ corresponding to the normalisation constant. The approach is further detailed by Peixoto[26], formulating the most succinct representation of the data as one with the minimum description length[36], $\Sigma$. Since the prior probabilities are nonparametric, the procedure also becomes parameter-free.

Choosing the model with the smallest description length $\Sigma$ is the means of balancing model complexity and goodness of fit[36]. We consider two candidate models throughout our experimental design: model $\mathcal{H}_a$, where layers are true descriptors corresponding to the conditional-probability weighted edges of links in genetic features in one layer and the frequency of sampling in another layer, and a null model $\mathcal{H}_b$ where both sets of edges are randomly interspersed across layers. Edge weights of both conditional probability and sampling frequency were resampled into the range space 0-1 and histograms reviewed to ensure comparable distributions for model fitting. The comparative magnitude of the description length of each model yields the following posterior odds ratio:

$$\Lambda = \frac{P(\{\theta\}_a|\{G_l\},\mathcal{H}_a)P(\mathcal{H}_a)}{P(\{\theta\}_b|\{G_l\},\mathcal{H}_b)P(\mathcal{H}_b)},$$

simplifying to

$$\Lambda = \exp(-\Delta\Sigma)\frac{P(\mathcal{H}_a)}{P(\mathcal{H}_b)}.$$

In this instance, $P(\{\theta\}|\{G_l\},\mathcal{H})$ is the posterior according to a given hypothesis $\mathcal{H}$, i.e., the true or null formulation. $P(\mathcal{H})$ is then the prior belief for hypothesis $\mathcal{H}$, and $\Delta\Sigma = \Sigma_a - \Sigma_b$ the difference in the model description length for these hypotheses. The description length of the true and null models can thus be formally compared. Where the description length of the true model (i.e., where sampling co-occurrence and conditional probability weights are correctly segregated by layer properties) is less than that of the null, then the model encoding sampling and conditional probability separately is preferred. Conversely, where the description length of the null is smaller, the layered formulation is shown to be superfluous, indicating the simpler, non-layered formulation should instead be preferred[21,26].

Next, we interrogated the structure of the graph with a nonparametric Bayesian stochastic block modelling approach. The result is a hierarchically organised community structure of nodes assigned into blocks that yields the most compact representation of the graph, as indexed by its minimum description length[36], $\Sigma$. Stochastic block models are described in extensive detail elsewhere[25,35,37]; their utility in neuroscience has been demonstrated and validated by multiple groups[21,22,25,37,38]. An evaluation of 275 empirical networks spanning a range of domains, including social, transport, information technological, and biological (including brain connectome data) has shown that networks whose diameter, $\varnothing$, is not large and random walk mixing times, $\tau$, are not slow are well suited to such modelling[37]. Z-scored with respect to the 275 surveyed networks, the parameters of our network were $\varnothing$ = -0.092 and $\tau$ = -0.11, well within the interval of well-modelled systems.

Having established the suitability of our approach, we proceeded to fit a stochastic block model to the genetic data, employing Markov Chain Monte Carlo (MCMC) to sample from the posterior of the estimated distribution. The MCMC procedure employed evidential equilibration by model entropy, the state of negative log-likelihood of the microcanonical stochastic block model, using Metropolis-Hasting acceptance-rejection sampling[21,22,39,40]. Following recommended practice[25,41,42], the chain was run with a stopping criterion of 100 iterations for a record-breaking event (i.e., an interval decrease to the description length), to ensure that equilibration was driven by changes in the entropy criterion. No burn-in is required since the MCMC proceeds from the initialised state generated by the stochastic block model itself. Bayesian model comparison based on minimum description length, in nats, was used to optimise correction by degree, nesting, and the choice of distribution (Gaussian or exponential) of the conditional probability edge weights. The centrality metrics of the inferred community structure were compared with one-way ANOVA.

## Stochastic block modelling of patient genetic signatures

The foregoing models reveal the community structure of the relations between genetic features, conditioning against linkages merely driven by sampling panel frequencies. We now proceed to model the community structure of the relations between individual patients shaped by their shared tumour genetic characteristics. The inferred structure is interpretable as a patient-level representation based on characteristic, signature genetic patterns. We hypothesized that this network representation would yield higher quality stratification of survival than either diagnostic labels or linear representations of genetic factors, demonstrating successful capture of tumour heterogeneity. We used a Sankey chart to visualize the links between known genetic mutations and the current best-practice diagnostic nomenclature[27], illustrating the diagnostic heterogeneity a stochastic block model representation could theoretically capture.

We created a dense graph with each patient, defined as a node, connected to every other by an undirected edge. Each edge was then independently weighted by the count of each genetic feature shared by the connected pair, resulting in a dense, fully connected graph with multiple binomial edge covariates spanning the full set of modelled tests. The number of edges is given as the number of nodes (patients) choose 2, $\binom{N}{2}$. We visualised the graph as a minimum spanning tree labelled by WHO CNS5 diagnosis or survival, enabling a qualitative impression of its expressive power in comparison with a non-graph linear model of the low-dimensional structure of the data based on principal component analysis (PCA).

We proceeded to fit and optimise a stochastic block model as outlined in the previous section, yielding a hierarchical community structure of patients. The z-scored $\oslash$ and τ parameters of our network were -0.067 and -0.098 respectively, again within the interval of well-modelled graphs. We then used Bayesian multinomial regression[43] to quantify the contribution of each genetic feature to each community. The multinomial regression was estimated with MCMC, employing a single chain running to 100,000 samples, a burn-in of 100,000 and thinning of 5, reporting the regression coefficient estimated with 95% Bayesian credibility interval.

### Survival modelling

To quantify the stratifying power of our network representation, we examined the prediction of survival, in days from the date of biopsy, for patients surveyed over at least 3 years. Date of biopsy was used as the index of onset in keeping with established practice in the field[44,45].

We sought to compare survival models based on i) our network genetic signatures, ii) patient diagnosis, and iii) the raw tumour genetic information used to fit the stochastic block model. We first constructed Cox's proportional hazard models, employing graph representational signatures, diagnosis, or raw tumour genetics across different models, with age and gender as nuisance covariates. We used a penaliser term of 0.1, and the Breslow baseline estimation method[46]. Model performance was evaluated by 5-fold cross-validation, relying on the median out-of-sample concordance index[46,47]. We extracted the survival function and hazard ratios of graph communities, diagnoses, and individual genetic domains for downstream comparison.

We augmented this analysis with a series of Bayesian logistic regression models[48,49], predicting survival at 12, 24 and 36 months, motivated by the widespread use of annual survival-based metrics[50,51]. These classification models replicated the inputs of the survival models, and were estimated with MCMC, employing 100,000 samples, a burn-in of 100,000 and thinning of 5. A series of prior shrinkage schemes were evaluated, including g, horseshoe, horseshoe+, ridge, lasso, and logt[48]. Model performance and goodness-of-fit were determined by pseudo-$R^2$ and the widely applicable information criterion (WAIC)[52], respectively.

The decision to evaluate the performance of network signatures against models of diagnosis or raw genetic features was driven by two factors: data requirements and favourable parameterisation. First, a systematic review of brain tumour survival models undertaken revealed that no previously published model incorporated the range of molecular data we had curated, studied different cohorts of the glioma landscape (e.g. just glioblastoma alone), and/or mandated additional data either not acquired during routine clinical care (e.g., full genome sequencing or proteomics), and/or necessitated multi-modal combination with medical imaging[53-70]. While these areas are undoubtedly interesting and add value to the field, our focus was to provide a means of forecasting survival with genetic data acquired in *routine clinical care* across the range of diagnoses available to us. Therefore, it was deemed appropriate to derive comparator models that would be tested against the graph-representations criterion on the original genetic data, and the WHO CNS5 diagnosis[27]. Second, it was important that our comparator models were comparable architecturally, so that any differences in model fidelity could be plausibly attributed to the quality of the representations, and not the hyperparameters/architectures that fit them. For this reason, it was judged appropriate to fit univariate models of diagnosis and *linear* multivariate models of genetics, but not *nonlinear* multivariate models of genetics. With all possible feature interactions here, the model parameter space rises to 3 628 800, which is clearly too large a space for a discriminative model supported by only 1323 patients. A non-linear model is therefore likely to overfit.

### Null models

We evaluated a series of nulls of the preceding models, created by randomly permuting edge features before following exactly the same modelling steps. Model comparison to the corresponding null by description length allows us to infer that the structure of a target model does not arise by chance. We additionally quantified the difference in the predictive performance of survival models based on the inferred community structure.

### Data and code availability

All code shall be made publicly available upon publication at https://github.com/high-dimensional. Trained model weights are available upon request. Patient data is not available for dissemination under the ethical framework that governs its use.

### Software

Analyses were predominantly performed within a Python (version 3.6.9) environment with the following software packages: graph-tool[42], GeoPy[71], gravis[72], hdbscan[73], lifeline[46], Matplotlib[74], NumPy[75], pandas[76], SciPy[77], seaborn[78], scikit-learn[47], statsmodels[79], and UMAP[80]. Bayesian logistic survival models were performed using MATLAB (version R2019a) with software package BayesReg[48], and multinomial logistic regression in R (version 4.1.3) using software package UPG[43].

### Compute

Analyses were performed on a 32-core Linux workstation with 128Gb of RAM and an NVIDIA 2080Ti GPU.

# Results

## Cohort

The 2021 WHO Classification of Tumours of the CNS (WHO CNS5) diagnoses across 4023 eligible patients included glioblastoma, IDH-wildtype (n=1713, 1015 male, 657 female, median age 60.52 years, IQR 16.55 (51.99-68.46)); astrocytoma, IDH-mutant (n=1186, 635 male, 495 female, median age 38.18 years, IQR 15.31 (31.20-46.51)); oligodendroglioma, IDH-mutant and 1p/19q codeleted (n=836, 424 male, 364 female, median age 44.36 years, IQR 19.29 (35.26-54.55)); and 'other glioma' tumour diagnoses (n=288, 151 male, 127 female, median age 25.32 years, IQR 29.82 (13.24-43.06)) (Table 1, Supplementary Figure 1).

| Full cohort | |
| --- | --- |
| Patient count | n=4023 |
| Age (years) | 51.72 (IQR 37.22 - 64.79) (100% available) |
| Sex | 2225 male, 1643 female (96% available) |
| Survival (days) | 979, IQR 324-2076 (33% available) |

| 2021 WHO Classification of Tumours of the CNS (WHO CNS5) | Glioblastoma, IDH-wildtype | Astrocytoma, IDH-mutant | Oligodendroglioma, IDH-mutant and 1p/19q-codeleted | Other Glioma |
| --- | --- | --- | --- | --- |
| Patient count | n=1713 | n=1186 | n=836 | n=288 |
| Age (years) | 60.52 (IQR 51.99-68.46) (100% available) | 38.18 (IQR 31.20-46.51) (100% available) | 44.36 (IQR 35.26-54.55) (100% available) | 25.32 (IQR 13.24-43.06) (100% available) |
| Sex | 1015 male, 657 female (98% available) | 635 male, 495 female (95% available) | 424 male, 364 female (94% available) | 151 male, 127 female (97% available) |
| Survival (days) | 365 (IQR 160-696) (38% available) | 1778 (IQR 1159-2518) (32% available) | 2198 (IQR 1528-3263) (28% available) | 2195 (IQR 1195-3485) (21% available) |

| Genetic Feature | Normal | Anomalous | Granularity of test result (where known) |
|---|---|---|---|
| IDH | Wildtype n=2001 (49.74%) (100% available) | Mutant n=2022 (50.26%) (100% available) | IDH1 G395A (n=360), IDH1 C394T (n=86), IDH1 C394G (n=53), IDH1 C394A (n=51), IDH1 G395T (n=13), IDH1 G394T (n=1), IDH2 G515A (n=11), IDH2 A514T (n=4), IDH2 A514G (n=3), IDH2 G515T (n=2) (29% available) |
| MGMT | Unmethylated n=2398 (59.61%) (100% available) | Methylated n=1625 (40.39%) (100% available) | 0-5% methylation (n=450), 5-10% methylation (n=536), 10-25% methylation (n=192), >25% methylation (n=447) (100% available) |
| EGFR | No amplification n=2885 (71.71%) (100% available) | Amplification n=1138 (28.29%) (100% available) | Low amplification (n=409), Moderate amplification (n=209), High amplification (n=520) (100% available) |
| 1p/19q | No codeletion n=2994 (74.42%) (100% available) | Codeletion n=1029 (25.58%) (100% available) | 1p/19q co-deletion (n=838), 19q deletion (n=191) (100% available) |
| Histone | Wildtype n=3962 (98.48%) (100% available) | Mutant n=61 (1.52%) (100% available) | Hist K27M (n=48), Hist G34R (n=13) (100% available) |
| TERT | Wildtype n=2096 (52.1%) (100% available) | Mutant n=1927 (47.9%) (100% available) | TERT C228T (n=993), TERT C250T (n=360) (70% available) |
| ATRX | Retained n=2009 (62.28%) (80% available) | Loss of expression n=1217 (37.72%) (80% available) | |
| BRAF | Wildtype n=3796 (94.36%) (100% available) | Mutant n=227 (5.64%) (100% available) | BRAF 1799 T>A (n=105), BRAF frameshift (n=5), BRAF Exon 16-9 (n=81), BRAF Exon 15-9 (n=21), BRAF Exon 16-11 (n=15) (100% available) |

Table 1 – Data: distributions of the total cohort, with WHO CNS5 diagnosis and with tumour genetic samples. Age and survival are both given as median with interquartile range, $25^{th}$ and $75^{th}$ percentiles.

In line with prior expectations, there were significant differences in age between all diagnoses (ANOVA p<0.0001, post-hoc Tukey honest significance tests all p<0.001), yielding an oldest to youngest order from i) glioblastoma, IDH-wildtype; ii) oligodendroglioma, IDH-mutant, and 1p/19q-codeleted; iii) astrocytoma, IDH-mutant; to iv) other glioma. There was an overall preponderance of men (p<0.0001), with modulation by specific diagnosis: more men were diagnosed with glioblastoma, IDH-wildtype (odds ratio 1.20, 95% confidence interval (CI) 1.067 to 1.35, p=0.002), but fewer with oligodendroglioma, IDH-mutant and 1p/19q-codeleted relative to the overall gender imbalance (odds ratio 0.85, 95% confidence interval (CI) 0.73 to 0.98, p=0.03). Overall median patient survival, in days from the date of surgery and tumour tissue sampling, was 979 days, IQR 1753 (324-2076).

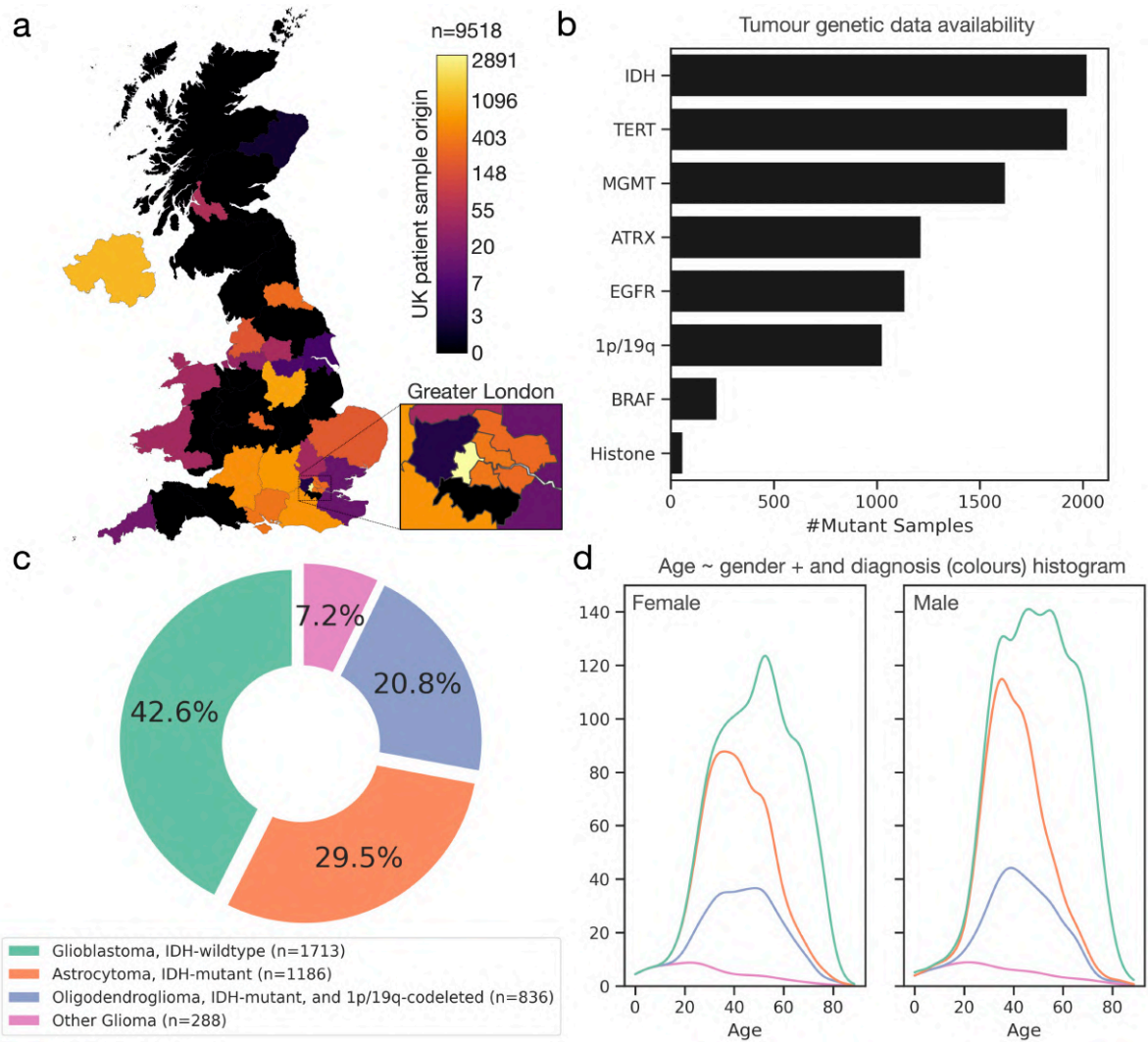

Figure 1 – Data: distributions by geography, tumour (epi)genetics, diagnoses, and demographics. a) Geographical distribution of all neuro-oncology patient data in the UK referred to our Division of Neuropathology between 2006-2020 for molecular diagnostics, in logarithmic axis per the colour bar. b) Number of mutant samples across the n=4023 glioma patient cohort. c) Distribution of WHO CNS5 diagnoses in cohort. d) Age kernel density estimators for male and female, subdivided to the diagnoses with corresponding colours as in panel c.

## Graph models of tumour genetic inter-relations

We used all available genetic data to create a comprehensive graph model with 37 tumour genetic features as nodes and 1332 non-zero directed edges weighted by Bayesian conditional probability of co-occurrence (Figure 2, Supplementary Figure 2). A one-way ANOVA of network centrality of tumour genetic features illustrated a significant difference in both conditional-probability weighted hub centrality ($p<0.0001$) and betweenness centrality ($p=0.026$) across features, indicating differing extents of inter-relatedness. Nodes with the greatest hub centrality—features linking to many others—were, in descending order, TERT, IDH, MGMT, 1p/19q, Histone, EGFR, ATRX and BRAF (Supplementary Figure 2). Of note, for some features, the specific genetic change—preservation of 1p/19q vs codeletion, for example—varied in hub centrality to a greater extent than changes across features. Genetic domains with the greatest betweenness centrality—features lying on the shortest path between others—were, in descending order, BRAF, EGFR, MGMT, IDH, Histone, ATRX, 1p/19q and TERT. A BRAF exon 16-11 mutation received the highest betweenness centrality within the BRAF category, the IDH1 G394T mutation received a far higher betweenness centrality than the remainder of the IDH features (despite being a relatively uncommon IDH mutant in our sample). Histone wildtype had a much lower betweenness centrality than anomalous histone features. The remaining centrality metrics did not vary significantly.

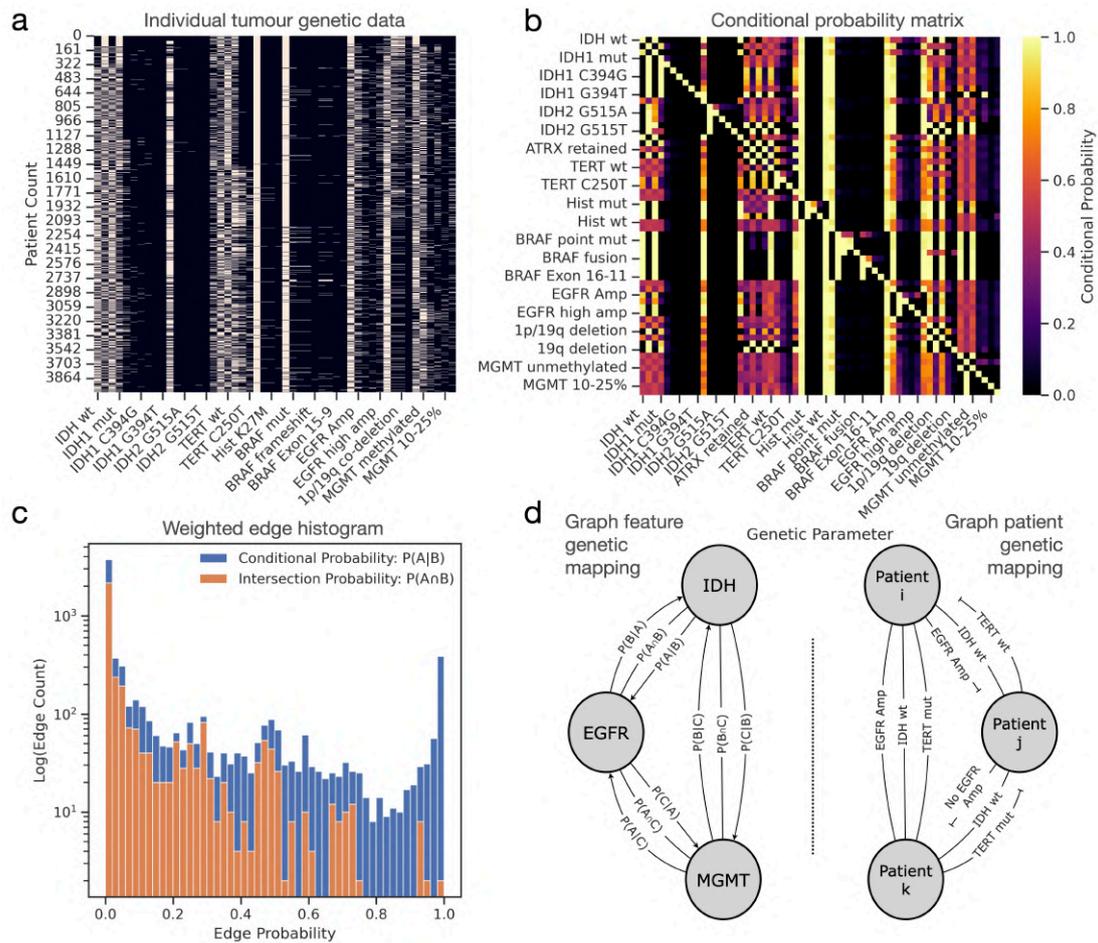

**Figure 2 – Method: graph modelling of brain tumours.** a) Rich genetic feature sets are extracted from patient histopathology report data (y-axis) and fractionated into individual genetic lesions (x-axis). The approach yields patient feature 'barcodes' corresponding to the complete molecular data available for a given patient's tumour. Note only a subset of features is labelled owing to visualization constraints. b) Heatmap of the conditional probability of one genetic feature given the presence of another, (P(A|B)), derived across the feature space, yielding an asymmetric adjacency matrix to be modelled as a directed Bayesian graph. In panels a and b, only a subset of features are labelled on the axes for visualization purposes. c) Histogram of edges in the Bayesian genetic network, with the number of edges present (y-axis, logged), and the corresponding probability assigned to the weighted edge (x-axis). The conditional weighted graph (blue bars) exhibits greater variation in inter-relatedness compared with the intersection between features (orange bars). d) Schematic illustrating the application of graph modelling for two purposes: i) *Graph feature genetic mapping*, where tumour genetics are modelled as nodes and their relations as probabilistically weighted, directed edges, and ii) *Graph patient genetic mapping*, where individual patients are nodes and edges are weighted by individual genetic features.

We fitted a layered, nested stochastic block model of all genetic features, with a multigraph of edges weighted, in separate layers[26], by conditional probability, and by sampling frequency. This approach allowed us to disentangle essential inter-relations between genetic features from the incidental effects of varied sampling. The model produced a structured graph representation, grouping features into hierarchically arranged "communities" with similar contributions to the overall graph structure. It should be stressed that communities can include mutually exclusive nodes: indeed, the presence of two mutually exclusive features within a community would plausibly signify that the two features have similar effects on the remaining graph.

The representation exhibited a hierarchical structure with 5 levels ($L_0$ to $L_4$) (Figure 3), organised into 37 ($L_0$), 8 ($L_1$), 3 ($L_2$), 2 ($L_3$) communities at each respective level, converging to a single block at $L_4$. Each level, $L_0$ to $L_4$, refers to a branch of the agglomerative hierarchical tree. The $L_1$ level revealed seven communities driven by genetic inter-relations as follows: i) EGFR amplification ± MGMT methylation ± 1p/19q deletion ± TERT C228T mutants; ii) IDH1 G395A mutants ± EGFR amplification; iii) IDH1 C394T ± IDH1 C394G ± IDH1 C394A mutants; iv) IDH2 G515A ± IDH1 G395T ± Histone G34R mutants; v) BRAF 1799 T>A ± BRAF exon 16-9 ± Histone K27M mutants; vi) IDH2 G515T ± IDH1 G394T ± IDH2 A514G ± BRAF frameshifts ± IDH2 A514T mutants, and vii) BRAF exon 15-9 and 16-11 mutants (Figure 3). The remaining community, primarily driven by sampling frequency, incorporated ATRX changes, IDH wildtype, unmethylated MGMT, BRAF, TERT, 1p/19q and histone wildtypes, and absence of EGFR amplification.

Bayesian model comparison ranging across hyperparameters identified the most plausible fit to be the layered, nested, degree-corrected model, with exponential weighting to the directional conditional probability edges (-1066.064 nats, with a posterior odds ratio of $e^{1508.3}$ favouring this fit incorporating conditional probability and sampling frequency separately over the random distribution null (Supplementary Figure 3). A non-layered stochastic block model with the sampling frequency layer ablated yielded a very similar structure, suggesting the absence of material sampling-related bias in the inferred patterns (Supplementary Figure 4). In contrast, the randomized null models both failed to derive a meaningful community structure and yielded far larger description lengths indicative of poor fit: randomized layered model = 1202.971 nats; randomized null model = 1522.296 nats (Supplementary Figure 5).

ANOVA of centrality metrics of the features within these communities identified a statistically significant difference in eigenvector centrality ($p<0.0001$), authority centrality ($p<0.0001$), hub centrality ($p<0.0001$), page rank ($p<0.0001$) and betweenness centrality ($p=0.01$) (Figure 3).

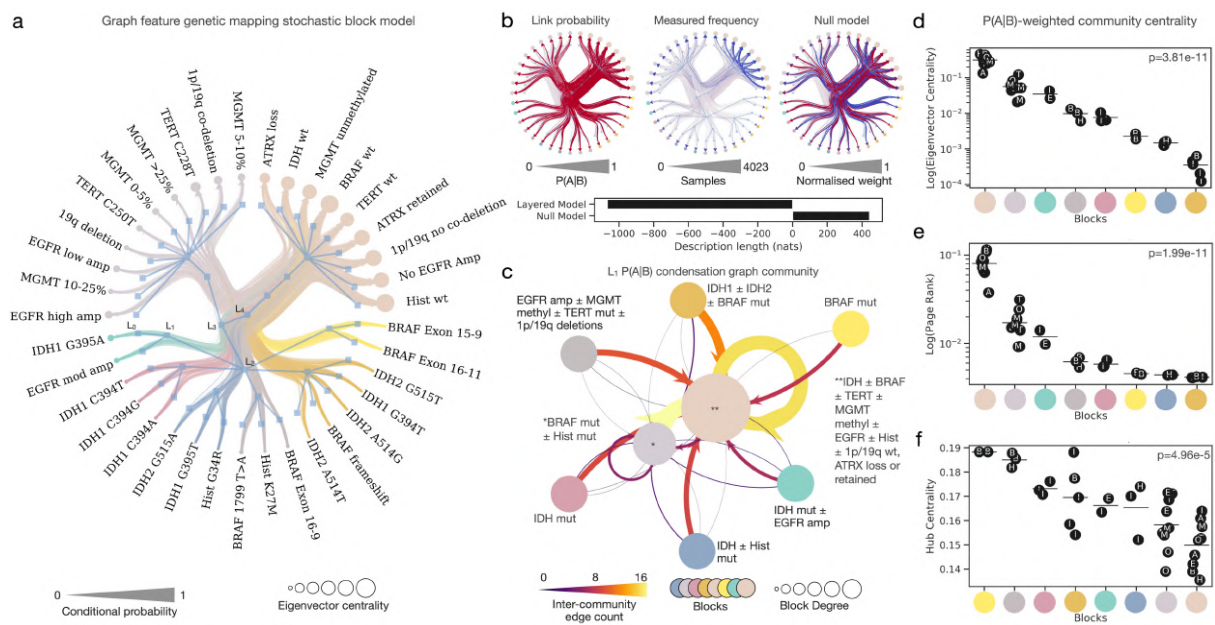

Figure 3 – Graph feature genetic mapping identifies characteristic genetic inter-relations. a) Radial graph of layered, nested, degree-corrected, and exponentially weighted stochastic block model revealing the community structure of tumour genetics and their influence upon overall network topology. Communities are colour coded by the first level community of the hierarchical community structure. Edges are sized according to their conditional probability. Nodes are sized according to their weighted-eigenvector centrality. Hierarchical levels are annotated from level 0 ($L_0$) to level 4 ($L_4$). b) Fits in accordance with link probability by conditional probability, measured frequency and the comparative null illustrate the description length of the layered model lower than the null, evidencing it a more suitable structure. c) Visualization of the first level hierarchy ($L_1$) with node colour as per that of panel a. Edge size and colour is proportional to the incidence of edges linking mutations between a given community. Node size is proportional to the degree of the corresponding community. d) There is a significant difference in weighted eigenvector centrality of tumour genetic factors ($p<0.0001$), e) page rank ($p<0.0001$) and f) hub centrality of tumour genetic factors when organized by stochastic block model community ($p<0.0001$). In panels d-f, block colour as per that of panel a, points are labelled by their corresponding abbreviation: A, ATRX; B, BRAF, E,

EGFR; H, Histone; I, IDH; M, MGMT, O, 1p/19q; T, TERT. Supplementary Figures 3-7 also accompany this plot with additional results.

Visualization of these inferred genetic inter-relations reproduced our current understanding of tumour genetics[27,81-87], but offered further insights into the interactions between genetic domains in generating the overall tumour genetic profile. We provide a downloadable and fully interactive graph representation of this molecular pathology landscape as Supplementary Figures 6 and 7 and encourage readers to use these interactive tools to guide further hypothesis testing in their research. A static image of coarse genetic domains is also shown in Supplementary Figure 2. We provide a breakdown below of the genetic inter-relations identified here that both affirm known findings and illuminate new relations, with quantitative metrics of the inter-relations directly retrievable from the supplementary figures appended.

## IDH

We confirm the association of IDH wildtype status with ATRX retention[81], often with TERT mutants (C228T and C250T in our dataset). In the presence of IDH wildtype, ATRX retention, and TERT mutation, a lesion was always 1p/19q wildtype. We demonstrate the propensity for at least low-level EGFR amplification in these IDH wildtype lesions[87,88]. We replicate the association of IDH mutants with ATRX loss in astrocytoma, and of TERT mutations, ATRX retention, and 1p/19q codeletions with oligodendrogliomas[83-86,88-90]. Within the finer IDH-mutant landscape, we demonstrate a strong association between IDH1 C394A mutants, TERT wildtype, no EGFR amplification, and ATRX loss in astrocytoma. IDH1 G394T and IDH2 A514G mutants exhibited the same characteristics, except that they were differentiated from each other by the presence of MGMT methylation in the 5-10% region in the former and not the latter[89,91,92]. IDH2 G515T mutants exhibited the previously reported[93] association with 1p/19q codeletions, as well as ATRX retention, no EGFR amplification, and unmethylated MGMT, in oligodendroglioma. The diversity of IDH mutations, interacting with other features, highlights the potential value of modelling tumour genetic data at a fine granularity.

## ATRX

Retained ATRX is confirmed to be associated with IDH wildtype in glioblastoma, as well as with the presence of IDH-mutants, TERT mutants and 1p/19q codeletions, in oligodendroglioma. ATRX loss is also associated with IDH mutants in astrocytoma[81-83,94-97]. We reveal a heterogeneity in the association of ATRX with different IDH mutants (see IDH section above). ATRX loss was indicative of non-amplified EGFR, largely in astrocytoma. Lastly, while ATRX loss was confirmed to be associated with preservation of 1p/19q, isolated 19q deletion was found to occasionally exist with ATRX loss in 19q-deleted astrocytoma[98]. Both histone G34R and K27M mutants could manifest ATRX loss or retention.

## EGFR

Non-amplified EGFR showed the expected association with IDH mutants (see section IDH), BRAF mutants, 1p/19q codeletions, and histone mutants (the latter largely in paediatric lesions). Any degree of EGFR amplification was associated with IDH wildtype, ATRX retention, variable TERT mutation status, and absence of a 1p/19q deletion, typically in glioblastomas[27,99]. We reveal the propensity for at least moderate, if not high-level, EGFR amplification to manifest with IDH wildtype glioblastoma.

## MGMT

Unmethylated or low-level (0-5%) methylated MGMT was associated with preservation of 1p/19q. 5-10% MGMT methylation, specifically, was found more in IDH1 G394T mutants. MGMT methylation levels were heterogenous across glioblastoma. Higher levels of MGMT methylation (>25%) indicated a lesion more likely to be ATRX-retained and histone wildtype.

## TERT

We confirm the known association of TERT wildtype with preserved 1p/19q and IDH mutants in astrocytoma, and of its mutants with IDH wildtype in glioblastoma, and IDH mutants in oligodendroglioma[100,101]. TERT wildtype was non-specific for both BRAF and histone wildtype and its mutants. Both TERT C228T and C250T promoter mutants were associated with IDH wildtype, preserved 1p/19q, histone wildtypes, and ATRX retention in glioblastoma cases. The mutually exclusive relationship between ATRX and TERT is confirmed[81,83]. Both TERT C228T and C250T mutants were also seen with IDH-mutants and 1p/19q codeletions.

## 1p/19q

We replicate the known exclusivity between 1p/19q codeletions and ATRX loss, where 1p/19q codeletion/ATRX retention is found in oligodendroglioma, but preserved 1p/19q and ATRX loss occurs in astrocytoma[81,95]. A 19q deletion alone was also associated with TERT wildtype and, interestingly, could also be seen with ATRX loss in astrocytomas[98]. We additionally reproduce the association of 1p/19q codeletion with EGFR amplification[102].

### Histone

Histone (K27M or G34R) altered tumours (typically diffuse hemispheric gliomas of paediatric/teenage and young adult demographic) were associated with IDH wildtype[27], as expected. But we also found that where either histone mutant was present, 1p/19q, BRAF, and TERT were wildtype, typically with no EGFR amplification. Histone K27M mutants were also less likely to exhibit MGMT methylation. Both K27M and G34R altered tumours could exhibit ATRX loss or retention, though ATRX loss was more likely in our histone G34R altered samples.

### Network signatures of tumour genetic heterogeneity

A Sankey chart visualizing the links between common genetic characteristics and diagnosis illustrates the marked genetic heterogeneity underlying established diagnostic categories (Figure 4). To determine if this structure can be revealed by graph models, we created a graph of the relations between patients defined by the similarities and differences of their tumour genetics. We began by creating a fully connected graph with 4023 patients as nodes and 8 090 253 edges. Each edge was weighted by 50 binary covariates indicating the status of all available genetic features at the finest available granularity. Visualising the graph as a minimum spanning tree labelled by WHO CNS5 diagnosis or survival showed distinct network patterns suggestive of greater differentiability between patients with systematically different outcomes than a principal component analysis representation of the same data.

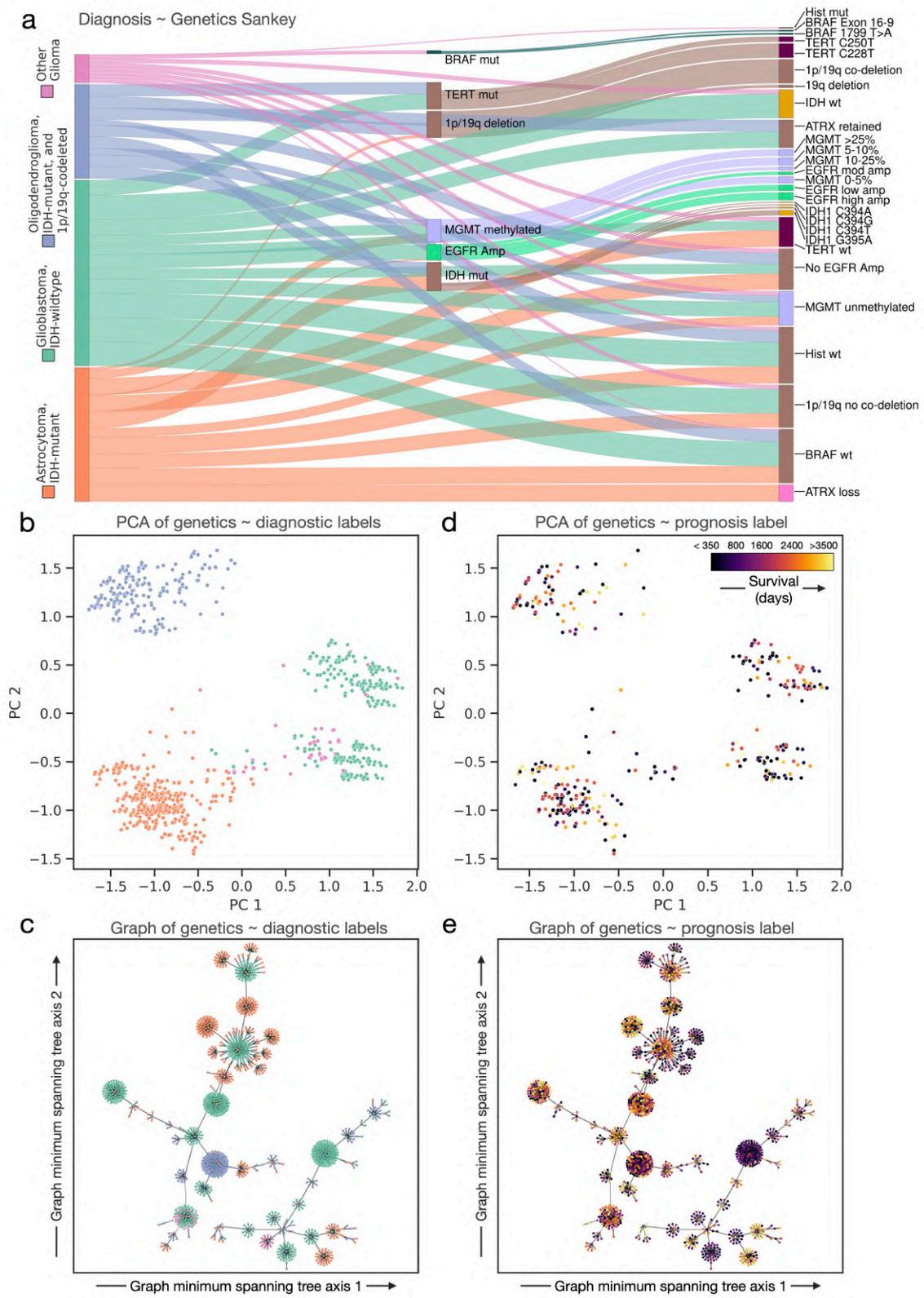

**Figure 4 – Tumour heterogeneity**. a) Sankey plot illustrating the variety of genetic features, under both coarse and finely granular descriptors, aligned to patient diagnosis. Only the most frequent links in our dataset are shown for readability. b) Principal component analysis of all tumour genetic data, which clusters individuals into patient groups reproducible of the diagnostic labels, colour coded as per the key. c) Minimum spanning tree of patients with edges weighted by the similarity between individual genetic tests appears to create a more richly structured representation of a tumour-genetic landscape, colour coded as in the key in panel c. d) Principal component analysis of all tumour genetic data with patient survival projected onto the plot illustrates a qualitatively poor representation of clusters of individuals with systematically better or worse survival. e) Minimum spanning tree of patients with edges weighted by the similarity between individual genetic tests with survival projected onto the plot illustrates a clearly superior segregation of individuals with better or worse prognosis, colour coded in the same way.

Indeed, a stochastic block model of the graph yielded a hierarchical community structure composed of 8 levels ($L_0$ to $L_7$), organised into 217 ($L_0$), 85 ($L_1$), 39 ($L_2$), 18 ($L_3$), 9 ($L_4$), 4 ($L_5$) and 2 ($L_6$) communities at each level respectively, converging to a single block at $L_7$ (Figure 5). As before, each level, $L_0$ to $L_7$, refers to a branch of the agglomerative hierarchical tree. There was clear evidence of MCMC model convergence, with a final model description length of 19 732 150 nats (Supplementary Figure 3). Each community exhibited characteristic patterns of demographic and genetic features, at the upper hierarchical levels reflecting the WHO CNS5 diagnosis (Supplementary Figure 8) but offering finer granularity below them. Examining median survival across these communities showed marked variation between communities with the same WHO CNS5 diagnosis (Supplementary Figure 9). For example, survival varied by 33.42%, from 324 to 454 days, across different clusters of patients with glioblastoma, IDH-wildtype. In contrast, the randomized null models both failed to derive a community structure which bore no resemblance to diagnosis and yielded a far larger description length indicative of poor fit (157 868 200 nats, i.e., an ~8-fold increase) (Supplementary Figure 10).

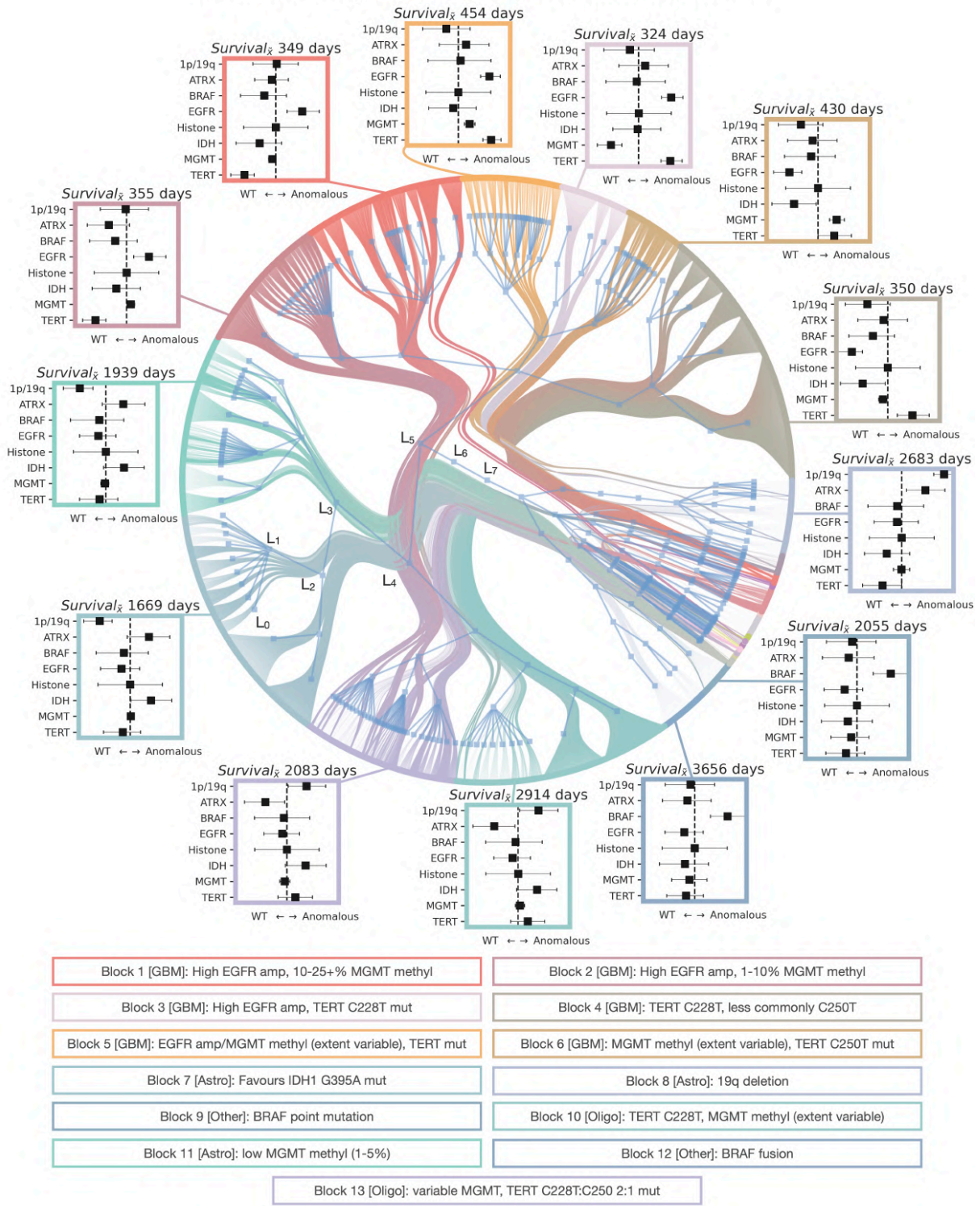

Figure 5 – Graph patient genetic mapping enables richer, more informative phenotyping. Radial graph of nested, degree-corrected, and multivariate binomially weighted stochastic block model revealing the community structure of patients and the genetics of their brain tumour. Hierarchical levels are annotated from level 0 ($L_0$) to level 7 ($L_7$). For visualisation purposes, communities are colour-coded by the second level blocks ($L_2$) of the hierarchical

community structure. Around the radial graph are the breakdown of median survival and box and whisker plots for the coefficients and 95% credible intervals of genetic loadings, where the coloured border of the plots depicts the corresponding community. All boxplots where the error-bar does not cross the vertical zero-line are significant, with features left of the vertical zero-line favouring the wildtype, and right of the zero-line favouring mutation. Supplementary Figures 3, 8-10 accompany this plot with additional results.

### Tumour genetic signatures of survival

To quantify the comparative prognostic power of the stochastic block model representation, we created separate survival models based on i) network signatures from the stochastic block model; ii) diagnosis; or iii) the raw tumour genetic and epigenetic data. Robust longitudinal survival data with >3 years follow-up was available for 1323 patients. Survival modelling with the network signatures revealed finer patient survival stratification, all modelled communities exhibiting statistically significant hazard ratios for either farer or poorer prognoses. Survival function curves based on these representations offered more closely individuated survival predictions, with specific hazard ratios for a given specific set of tumour genetic features (Figure 6 and Supplementary Figure 9). For example, two distinct communities, blocks 1 and 6, with the same diagnosis—glioblastoma, IDH-wildtype—yielded rather different hazard ratios: 2.76 vs 2.27, the former more likely to exhibit MGMT methylation and high EGFR amplification than the latter (Figure 6 and Supplementary Figure 9).

Survival models based on WHO CNS5 diagnosis achieved much cruder stratification, only distinguishing glioblastoma, IDH-wildtype (HR of 3.00 (median survival 365 days, IQR 536 (160-696)) from all others: astrocytoma, IDH-mutant (HR 0.55, median survival 1778 days, IQR 1358 (1159-2518)), oligodendroglioma, IDH-mutant and 1p/19q codeleted (HR 0.43, median survival 2198 days, IQR 1735 (1528-3263)) and the other gliomas (HR 0.63, median survival 2195 days, IQR 2290 (1195-3485)) (Figure 6, Table 1). Survival models based on the source tumour genetic data revealed comparatively few significant predictive features: the 95% confidence intervals of histone, ATRX and MGMT methylation HR all crossed 1, EGFR amplification and TERT mutants were significantly associated with poorer prognosis (HR 1.64 and 1.23, respectively). 1p/19q deletion and IDH mutations were both significantly associated with a better prognosis (HRs 0.54, and 0.39, respectively) (Figure 6 and Supplementary Figure 9).

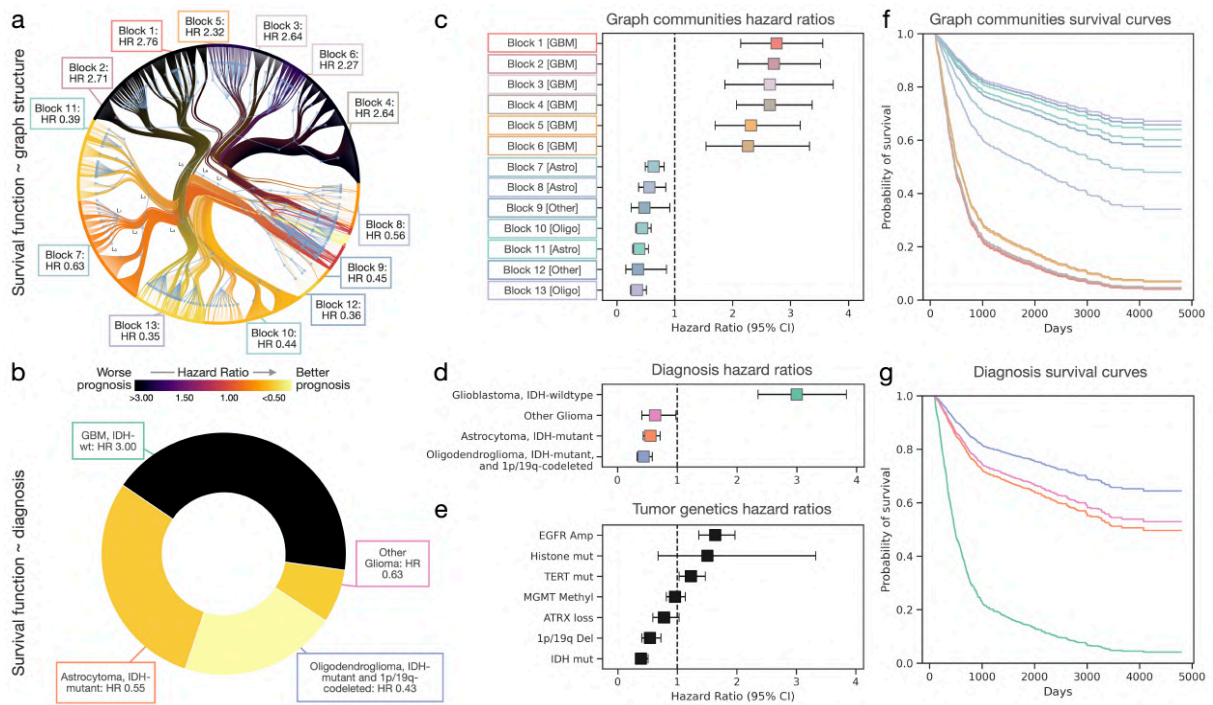

Figure 6 – Graph patient genetic mapping offers higher fidelity prognosis. a) Radial graph of a nested, degree-corrected, and binomially weighted stochastic block model revealing the community structure of patients based on tumour genetics (as also shown in Figure 5 and Supplementary Figures 8-9). Communities are colour coded by the hazard ratio of the survival model of the second level blocks ($L_2$). Note only the minimum spanning tree of the graph is shown, owing to visualisation constraints. b) Pie chart of brain tumour diagnoses colour-coded by the hazard ratio of the survival model with the diagnostic label. In a-b), darker colours convey a poorer prognosis (hazard ratio > 1), and conversely lighter colours a more favourable one (hazard ratio < 1). c) Box and whisker plot illustrating the hazard ratios with 95% confidence interval of the second level blocks of the stochastic block model community structure. d) Box and whisker plot illustrating the hazard ratios with 95% confidence interval of the tumour diagnoses, illustrating only a crude discrimination of glioblastoma, IDH-wildtype from the remainder. e) Box and whisker plot illustrating the hazard ratios with 95% confidence interval of the raw tumour genetics. f) Survival plot of the second level blocks of the stochastic block model community structure illustrates a rich variation in survival, colour-coded by the blocks in both panels a, e) and on Figure 5. g) Survival plot of the tumour diagnoses shows coarser forecasting of patient prognosis, colour-coded by the diagnoses of panels b. In panels c-e), all points where the whiskers do not cross the vertical line at 1 are statistically significant. Supplementary Figures 9-11 also accompanies this plot with additional results.

We conducted formal model comparison to determine whether network signatures, diagnosis, or raw (epi)genetic data (both inherently diagnostic—e.g., IDH status—and supplementary variables—e.g., MGMT methylation array) offered superior fidelity in forecasting survival. In keeping with established practice, models were statistically compared with $R^2$ and the widely applicable information criterion (WAIC), inferring the best model to be the one with the lowest WAIC. We did so with all plausibly expressive levels of the graph hierarchy ($L_1$ to $L_4$ agglomerative community blocks – see Figure 5), and with both continuous regression models (Cox's proportional Hazard), and Bayesian logits for 12-, 24- and 36-month survival (Figure 7).

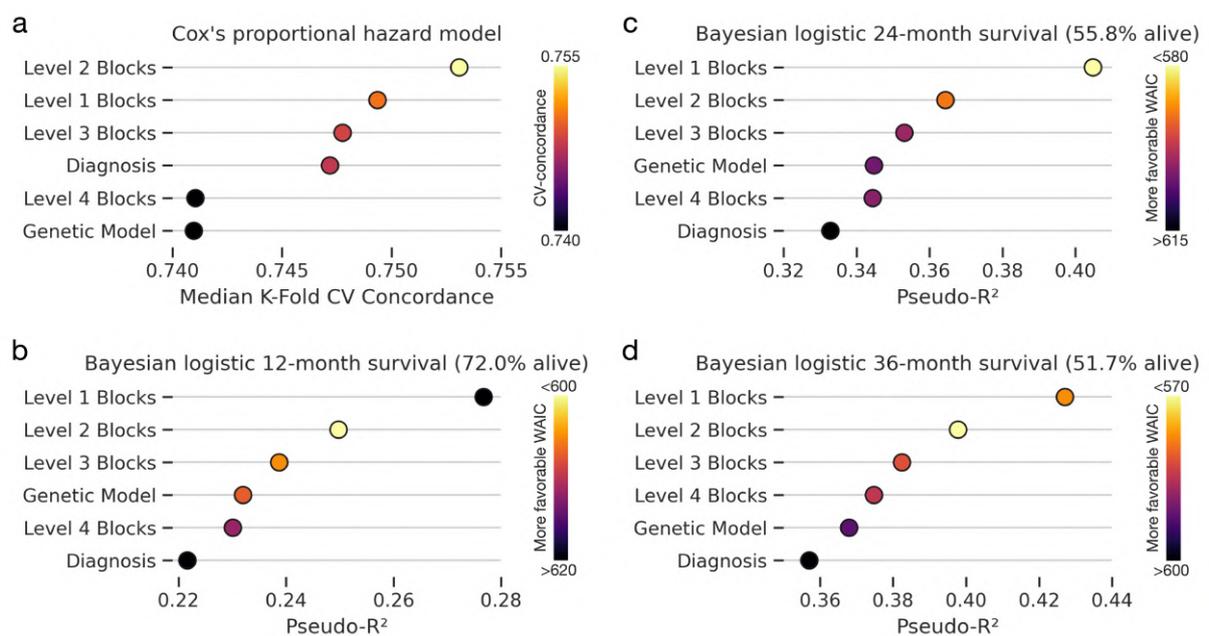

Figure 7 – Network signatures forecast survival better than WHO CNS5 diagnosis or raw genetic and epigenetic features. a) Model predictive performance evaluated by cross-validated concordance index of Cox's proportional hazard model shows network signatures outperform both models of diagnosis and the original genetic information in forecasting survival. b-d) Model predictive performance evaluated by pseudo-$R^2$ and WAIC of Bayesian logistic regression survival models for 12-month (panel b), 24-month (panel c) and 36-month (panel d) survival shows network signatures outperform both models of diagnosis and the original genetic information in forecasting survival, and with more favourable fits by WAIC (lower is better). Supplementary Figure 11 accompanies this plot with additional results.

Network signatures achieved the best out-of-sample predictive performance on Cox's proportional hazard modelling. Median cross-validated concordance, in decreasing order of performance, was 0.753 with $L_2$ graph blocks, 0.749 with $L_1$ graph blocks, 0.748 with $L_3$ graph blocks, 0.741 with diagnosis, 0.741 with $L_4$ graph blocks, and 0.740 with raw tumour (epi)genetics.

Network signatures also achieved the best out-of-sample predictive performance on Bayesian year-discretized survival predictions. For 12-month survival, where 72.0% of the cohort remained alive, the best performing models in descending order of $R^2$ were: $L_1$ blocks ($R^2$ 0.277, WAIC 617), $L_2$ blocks ($R^2$ 0.250, WAIC 610), $L_3$ blocks ($R^2$ 0.239, WAIC 611), raw tumour (epi)genetics ($R^2$ 0.232, WAIC 612), $L_4$ blocks ($R^2$ 0.230, WAIC 614) and diagnosis ($R^2$ 0.221, WAIC 617).

For 24-month survival, where 55.8% of the cohort remained alive, the best performing models in descending order of $R^2$ were: $L_1$ blocks ($R^2$ 0.405, WAIC 587), $L_2$ blocks ($R^2$ 0.364, WAIC 594), $L_3$ blocks ($R^2$ 0.351, WAIC 601), raw tumour (epi)genetics ($R^2$ 0.345, WAIC 604), $L_4$ blocks ($R^2$ 0.344, WAIC 602) and diagnosis ($R^2$ 0.333, WAIC 611).

For 36-month survival, where 51.7% of the cohort remained alive, the best performing models in descending order of $R^2$ were: $L_1$ blocks ($R^2$ 0.423, WAIC 577), $L_2$ blocks ($R^2$ 0.400, WAIC 571), $L_3$ blocks ($R^2$ 0.382, WAIC 581), $L_4$ blocks ($R^2$ 0.375, WAIC 583), raw tumour (epi)genetics ($R^2$ 0.368, WAIC 589), and diagnosis ($R^2$ 0.357, WAIC 595).

In contrast, survival models using randomized null graph models failed to derive any meaningful survival prediction, nor with any community segregation (Supplementary Figure 10), offering chance accuracy: CPH c-index 0.534; 12-, 24- and 36-month survival $R^2$ 0.008 or lower (Supplementary Figure 11).

## Discussion

We have developed a comprehensive framework, founded on Bayesian non-parametric models of the community structure of graphs, for extracting interactive biological patterns from routinely acquired high-dimensional brain tumour genetic data, modelling relations not only between individual genetic features, but also between individual patients, with large-scale, representative, fully-inclusive international data acquired prospectively over a 14-year period. Our framework has two aims: first, to reveal systematic genetic inter-relations potentially material to the pathogenesis of brain tumours over and above individual genetic contributions, thereby catalysing mechanistic hypothesis generation and therapeutic innovation, and second, to enable higher fidelity, more closely individuated patient stratification, with potential prognostic and prescriptive utility. Our approach not only successfully identifies known genetic inter-relations but reveals new ones, and not only replicates the WHO CNS5 diagnosis but provides a hierarchical patient stratification capable of predicting survival with higher individual-level fidelity than either diagnosis or simple linear models of the raw genetic and epigenetic features. Overall, these findings overwhelmingly support the value in applied network science in neuro-oncology[18].

### The demographic structure of brain tumour genetics

We identify striking heterogeneities in the demographics of genetically defined brain tumours and their subtypes in our dataset of operable patients. In line with the literature[3,27], patients with glioblastoma, IDH-wildtype were the oldest, followed in descending order of age by oligodendroglioma, IDH-mutant and 1p/19q codeleted, astrocytoma, IDH-mutant, and the remaining other gliomas (including BRAF mutant lesions characteristic of children and young adults). Overall, men were more prevalent than women in this large multi-site glioma sample, but significantly more so in glioblastoma, IDH wildtype, than the other tumours. Conversely, women were significantly more likely to be diagnosed with an oligodendroglioma, IDH-mutant and 1p/19q codeleted, when explicitly controlling for the cohort gender imbalance.

### The value of a network approach

Our analysis attests to the value of a graph modelling[15-18,22,25,26,37,41,103] in eliciting rich phenotypic information underpinning the genetic heterogeneity of brain tumours. We have shown that graph analysis can reveal hierarchical communities of tumour genetic features sharing similar patterns of inter-relatedness and influence upon an overall tumour genetic structure that plausibly have mechanistic implications for the manifestation of brain tumours. Such communities are potential targets for more detailed examination and should be investigated across future research.

Moreover, we illustrate how graph analysis provides not only a representation[35] of tumour genetics, but also of patients themselves across the tumour genetics landscape. Such a process automatically recovers the diagnostic labels with ease yet offers finer a granularity of patient subpopulations determined by specific, signature constellations of inter-related tumour genetics. The hierarchical nature of the representation provides a flexible means of parameterizing tumour genetics at a granularity optimised for the downstream task—e.g., predicting survival—and volume of available data. Where the volume of data is low, a coarse representation derived from upper levels of the hierarchical structure would be appropriate; where it is high, a finer representation becomes statistically tractable. Generative stochastic block models provide formal support for these representations, relying on a formal equivalence between compression and inference in the specific setting[104].

Critically, there is no theoretical constraint on the size of the models, only the impact of practical constraints such as data and compute that need be examined empirically. Future modelling could include other feature sets, such as more comprehensive genomics, exomics, or with features of *intra*-tumoural heterogeneity such as variant allele frequencies, sample error or purity. The algorithmic approach has been successfully applied to graphs of 3.38 million nodes (>840 times more than ours)[37], leaving plenty of room for expansion. Note that such community structure as is discernible in the current models suggests the domain is eminently suited to stochastic block modelling, moreover allied research suggests biological networks (as is the case here) are especially well suited to the methodological approach[37].

## Graph feature genetic mapping

Stochastic block models of the probabilities of one genetic feature conditioned on another yield a comprehensive hierarchical representation of tumour genetics shaped by the conditional relations between genetic features. This representation inevitably reflects well-known relations such as the mutual exclusivity of 1p/19q codeletion and ATRX loss, captured within large communities. But it also reveals hitherto unrecognized patterns within smaller, more specific communities, such as clustering of IDH1 G395A with moderate degrees of EGFR amplification, histone G34R with IDH1 G395T and IDH2 G515A mutants, and BRAF frameshifts with several IDH2 mutants (G515T, G394T, A514G), into community groupings that yield similar effects on the remaining graph genetic landscape, in spite of how many of these aforementioned mutations are mutually exclusive (including IDH mutants with EGFR amplification; histone-altered with IDH mutants). Explaining these, amongst other newly identified patterns, is the task of future research.

The genetic communities we have revealed demonstrate varying magnitudes of network centrality. For eigenvector and page-rank centrality—both approximate measures of a node's 'influence' in a network—those communities most 'influential', in descending order were: i) IDH ± BRAF ± TERT ± MGMT ± EGFR ± Hist ± 1p/19q wildtypes, non-specific ATRX loss/ret; ii) EGFR amplifications ± MGM methylation ± TERT C250T mutants ± 1p/19q deletions; iii) IDH mutants ± EGFR amplifications; iv) BRAF ± hist mutants; v) IDH mutants; vi) BRAF mutants; vii) IDH ± histone mutants; and vii) IDH1 ± IDH2 ± BRAF mutants. This would make plausible biological sense, given TERT mutation status can be mutant or wildtype across the glioblastomas, mutant in oligodendroglioma and typically wildtype in astrocytoma, therefore the range of possible TERT alterations cover a large proportion of tumour diagnoses[27,83,84,100,101]. In addition, BRAF alterations are essentially indicative of a set of specific tumour diagnoses[27,81,105-110], with relatively little scope for mutation amongst other diagnoses. Some of these patterns are expected—for instance, the presence of an IDH wildtype raises the probability for a glioblastoma with accompanying molecular pathology panel to follow[27,84,111]— but relations particular to the specific type of genetic lesion are not. We should note the marked variations in the hub-centrality of genetic features: whereas BRAF, followed by histone and IDH mutants demonstrated greater hub-ness, i.e., those features which link to many other possible features. This finding exemplifies the problem of tumour heterogeneity[5,8-10,112-115]: some features can link to a vast array of alternative molecular features[6], and might only be understood through a graph.

### Graph patient genetic mapping

Generating a stochastic block model of tumours defined by their genetic features reveals a complex hierarchical community structure reflecting patterns of genetic inter-relations varying systematically across patients. The model not only recovers the tumour diagnosis but provides a multi-level stratification of patients exhibiting different tumour genetic signatures.

We have shown that a finer description based on the more distal levels of the graph yields better predictive performance to their individual survival than WHO CNS5 diagnoses. The subcommunities of patients diagnosed with glioblastoma, IDH-wildtype provide a striking example. Here median prognosis systematically varies from 324 to 454 days. Specifically, the glioblastoma, IDH-wildtype cohort segregated into communities of progressively poorer prognoses, as follows: i) high EGFR amplification with 10-25+% MGMT methylation (HR 2.76); ii) high EGFR amplification with 1-10% MGMT methylation (HR 2.32); iii) high EGFR amplification with a TERT (C228T) mutation (HR 2.64); iv) TERT mutations (usually C228T, less commonly C250T), but non-amplified EGFR and unmethylated MGMT (HR 2.64); v) EGFR amplification, MGMT methylation (but with variable extent) and TERT mutations (HR 2.32); and iv) variable MGMT methylation with TERT C250T mutations (HR 2.27). That prognostication is part-explained by the presence or absence of genetic features in the tumour landscape has been raised elsewhere[81,84,116-118], but is particularly amplified in our findings where a detailed set of inter-relating features conveys greater survival fidelity over singular molecular profiles. We stress that while a median survival difference of 130 days might not seem substantial at the population level, it may be of great significance at the individual level, especially given how poor survival in glioblastoma is: 1 in 4 surviving beyond 2 years[119].

It should be noted that the utility of the patient-level representations we have derived here is likely to be closer to the floor than the ceiling of possibility. The traditional small data regimes that dominate the field typically enforce the use of univariate or low-dimensional linear multivariate models constitutionally blind to complex interactive effects. Such models yield comparatively few significant features, necessarily selected on their linear effects, that filter through to clinical use. The large – inevitably clinical – corpora as then accumulate are enriched in linear (though not necessarily purely linear) effects. A more expressive modelling framework, capable of capturing complex interactions, justifies casting the net more widely – in terms of case and feature numbers – but needs large-scale data of the right kind to substantiate. In developing such a framework, based on expressive mathematical models well-grounded in Bayesian inference and graph theory[35-37,104], we hope to stimulate wider recognition of the possibility of identifying potentially valuable factors concealed by their non-linearities.

## Highlighting genetic interactions

The vast majority of the genetic features studied here have been associated with a specific diagnosis and/or a particular prognosis[115], explaining their inclusion in routine clinical investigation. Nonetheless, the striking multiplicity of features demonstrates the natural complexity of oncogenesis[23,24]. It is from this premise that we argue for the value of the graph approach presented here[15,16,18]—which definitionally incorporates interactions between multiple features—as a means of illuminating disease processes, on which future treatment innovation inevitably depends.

Survival prediction illustrates the potential value of our approach. Take the following genetic features, all familiar in isolation: i) IDH – its wildtype form now signifies a tumour to be a glioblastoma[27,115] and associated with a poorer prognosis; ii) MGMT – greater degrees of methylation are associated with altered responsiveness to Temozolomide, and a fairer prognosis (although likely in part due to treatment allocation)[29,119-122]; iii) EGFR – greater amplification is associated with poorer outcomes[123]; and iv) TERT promotor mutants - telomere extension is thought essential to key neoplastic mechanisms[115,124]. All these features are individually prognostic to some degree, but their interactions cast further light, segregating patients into intersectional subpopulations whose prognosis varies substantially and systematically with the specific pattern of interaction (Figure 5, Figure 6, Supplementary Figures 8-9). It is not the case that simply more mutations equate to poorer prognosis, but rather specific sets of interactions dictate them, supporting the notion that sets of features are prognostic[81], rather than single factors taken in isolation. For instance, although isolated TERT mutants carry a poor prognosis (see Figure 6, panel e, and other studies[116,118]), a TERT wildtype paired with EGFR amplification and MGMT methylation yielded poorer prognoses than many other tumour genetic communities, including many of those exhibiting TERT promotor mutants.

## Enhancing individual-level prognosis

Network signatures of patient brain tumour genetic communities predict survival with greater fidelity than coarse diagnostic labels[27]. Predictive performance appeared competitive with comparable, dedicated genetic analyses performed by others[65,125,126]. For example, Chen et al., reported a survival model with C-index fidelity of 0.818[65], slightly higher than what we detail here, though is applicable to low grade gliomas only, whereas ours evaluates places no such inclusion criteria. Similarly, Yousefi et al., provide deep survival models for low- and high-grade lesions, reporting a c-index between 0.75-0.84[126], but these models require comprehensive genomic data rarely available as part of routine clinical care. We suggest that the inclusivity of our framework, and its dependence only on routinely acquired genetic data, allows us to cast the net more widely in pursuing associations with potential clinical value. Moreover, we show here that whereas survival modelling by diagnosis is primarily driven by the distinction between IDH wildtype glioblastoma and other diagnoses, the graph community structure offers a far more finely stratified result. Glioblastoma subpopulations faring better or worse hinged on specific genetic traits, with similarly varied survivability across more favourable diagnoses (Figure 6). It is intriguing that linear survival models constructed with the same tumour genetic data used to fit the graph community structure performed no better than diagnosis-based models. That the graph representation provides greater predictive power illustrates the potential value of harnessing the complex high-dimensional inter-relationships between tumour genetic features, and ought to stimulate further investigation.

Note that the superiority of network signatures was evident not only in Cox's proportional hazard modelling, but also in annually discretized classification within a Bayesian inferential framework. These models demonstrated more favourable goodness-of-fit by WAIC, indicating the superiority is not trivially explained by model overparameterization but by a better representation.

## Study limitations

We sought to reveal the nature and prognostic value of modelling the inter-relationships between tumour genetic features acquired in the context of routine clinical care. The computational complexity of the task mandates the assembly of a large-scale, fully inclusive set of data. Such a set inevitably requires accumulation of data over long periods, covering substantial changes in investigational and diagnostic practice[27]. We therefore adopted a careful, multi-step approach for appropriate handling of data missingness that rendered 4023 of 9518 patients prospectively curated from 2006 to 2020 eligible for inclusion. Our objective, however, is not to provide a definitive representation of tumour genetics, but to demonstrate a suitable approach to drawing intelligence from tumour genetic data in a manner sensitive to its complex interactions. For survival modelling, while we included the demographic features of age and sex, we could not include performance index or other clinical characteristics owing to their lack of availability. Naturally, where such data is available it ought to be modelled, and its value quantified through the kind of model comparison we perform here.

## Conclusion

Graph models of brain tumour genetics illuminate the landscape of tumour heterogeneity and enable better prognosis of survival than either diagnosis or models of individual genetic features. They offer a principled means of deriving rich phenotypic representations, with the finer descriptive granularity on which greater personalisation of care inevitably depends. Translation of such an approach to the clinical frontline may offer opportunity for better and more patient-focussed care.

# Brain tumour genetic network signatures of survival

# Supplementary material

James K. Ruffle FRCR MSc[1], Samia Mohinta MSc[1], Guilherme Pombo MSc[1], Robert Gray PhD[1], Valeriya Kopanitsa MBBS BSc[1], Faith Lee BSc[1], Sebastian Brandner MD FRCPath[2], Harpreet Hyare FRCR PhD[1], and Parashkev Nachev FRCP PhD[1]

[1]*Queen Square Institute of Neurology, University College London, London WC1N 3BG, UK*
[2]*Division of Neuropathology and Department of Neurodegenerative Disease, Queen Square Institute of Neurology, University College London, London WC1N 3BG, UK*

Running title:
Brain tumour genetic network signatures of survival

Correspondence to:
Dr James K Ruffle
Email: j.ruffle@ucl.ac.uk
Address: Institute of Neurology, UCL, London WC1N 3BG, UK

Correspondence may also be addressed to:
Professor Parashkev Nachev
Email: p.nachev@ucl.ac.uk
Address: Institute of Neurology, UCL, London WC1N 3BG, UK

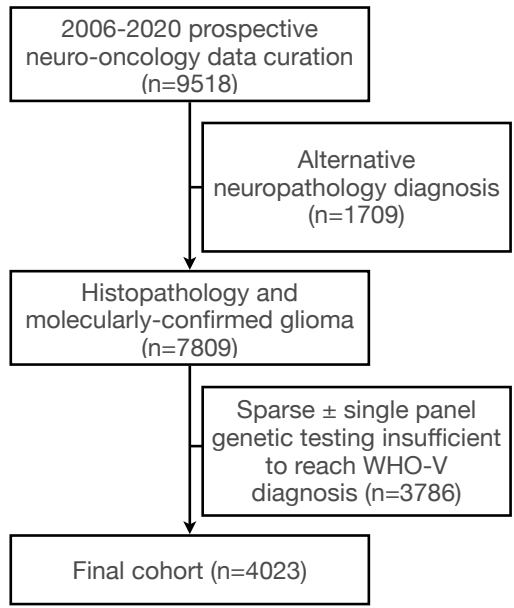

Supplementary Figure 1 - Study flow chart.

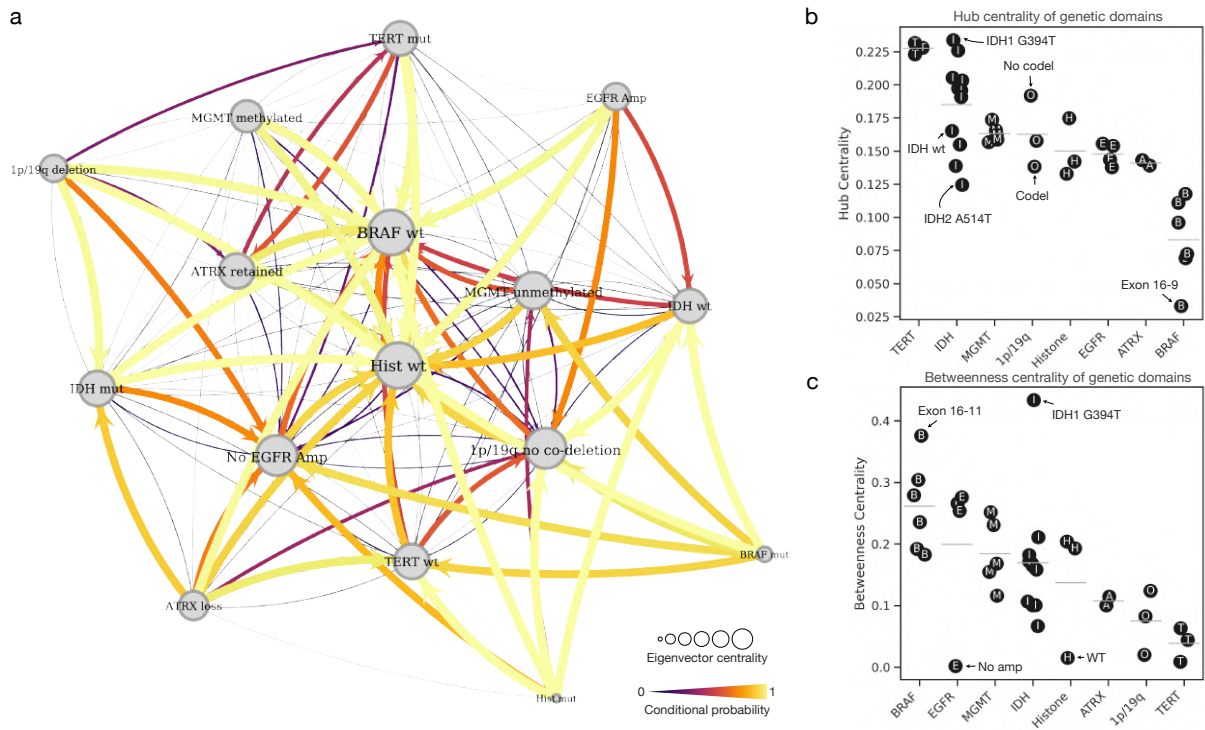

**Supplementary Figure 2– Graph models of course tumour genetic domains.** a) Spring-block layout of course genetic features - i.e., mutant or wild-type - where the size and colour of edges are proportional to the directional conditional probability, and node size is proportional to the weighted eigenvector centrality. b) There is a significant difference in weighted hub-centrality of tumour genetic factors when organized by genetic domain (p<0.0001). c) There is a significant difference in weighted betweenness-centrality of tumour genetic factors when organized by genetic domain (p<0.0001). In panels b-c) points are labelled by their corresponding abbreviation below and outliers are also annotated: A, ATRX; B, BRAF, E, EGFR; H, Histone; I, IDH; M, MGMT, O, 1p/19q; T, TERT.

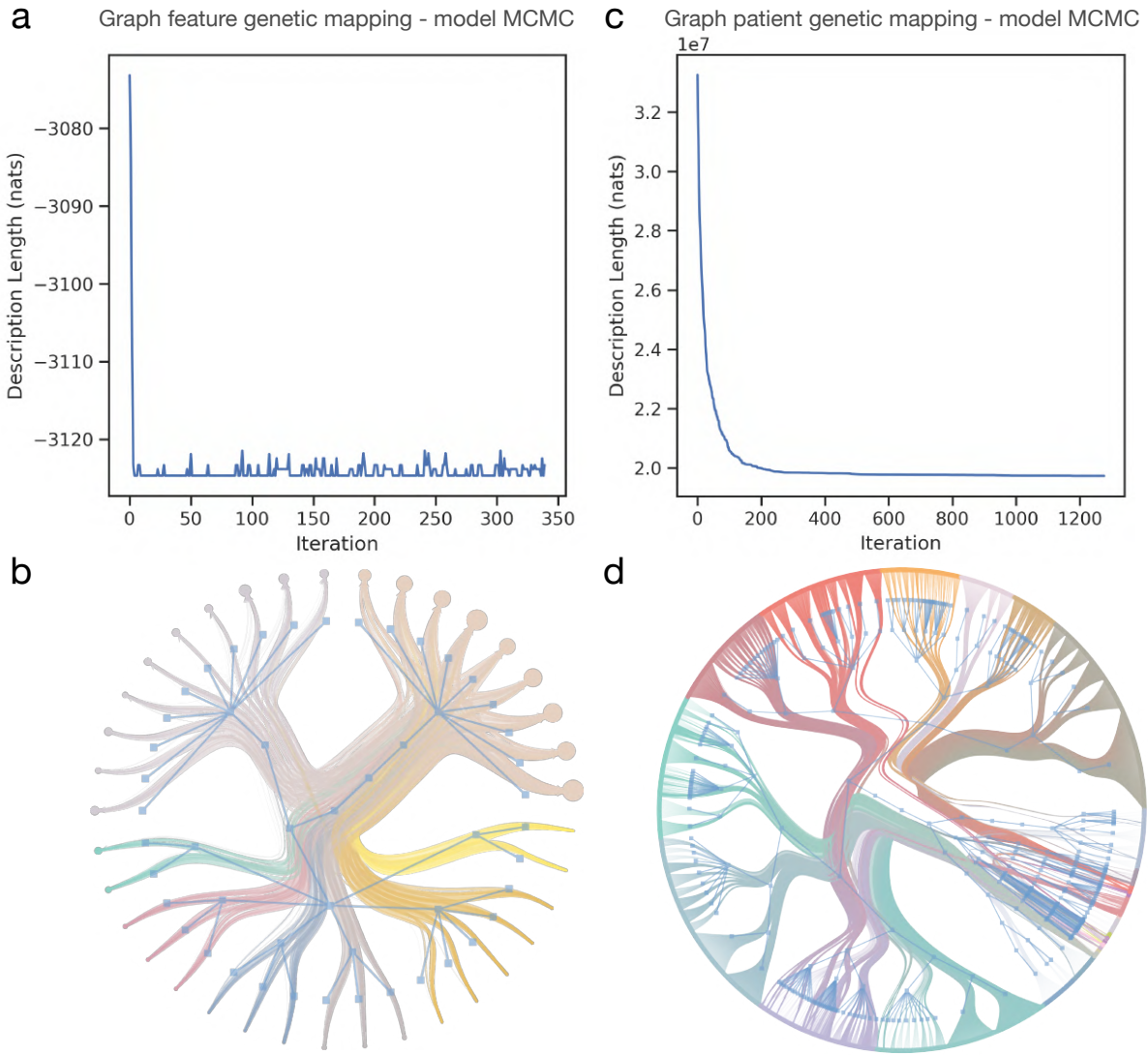

Supplementary Figure 3 – Evidence of model convergence. a) Iterative reduction in the description length of the layered graph feature genetic mapping nested stochastic block model with MCMC, which evidences convergence to the most plausible fit as shown in panel b). c) Iterative reduction in the description length of the graph patient genetic mapping model with MCMC, which evidences convergence to the most plausible fit as shown in panel d).

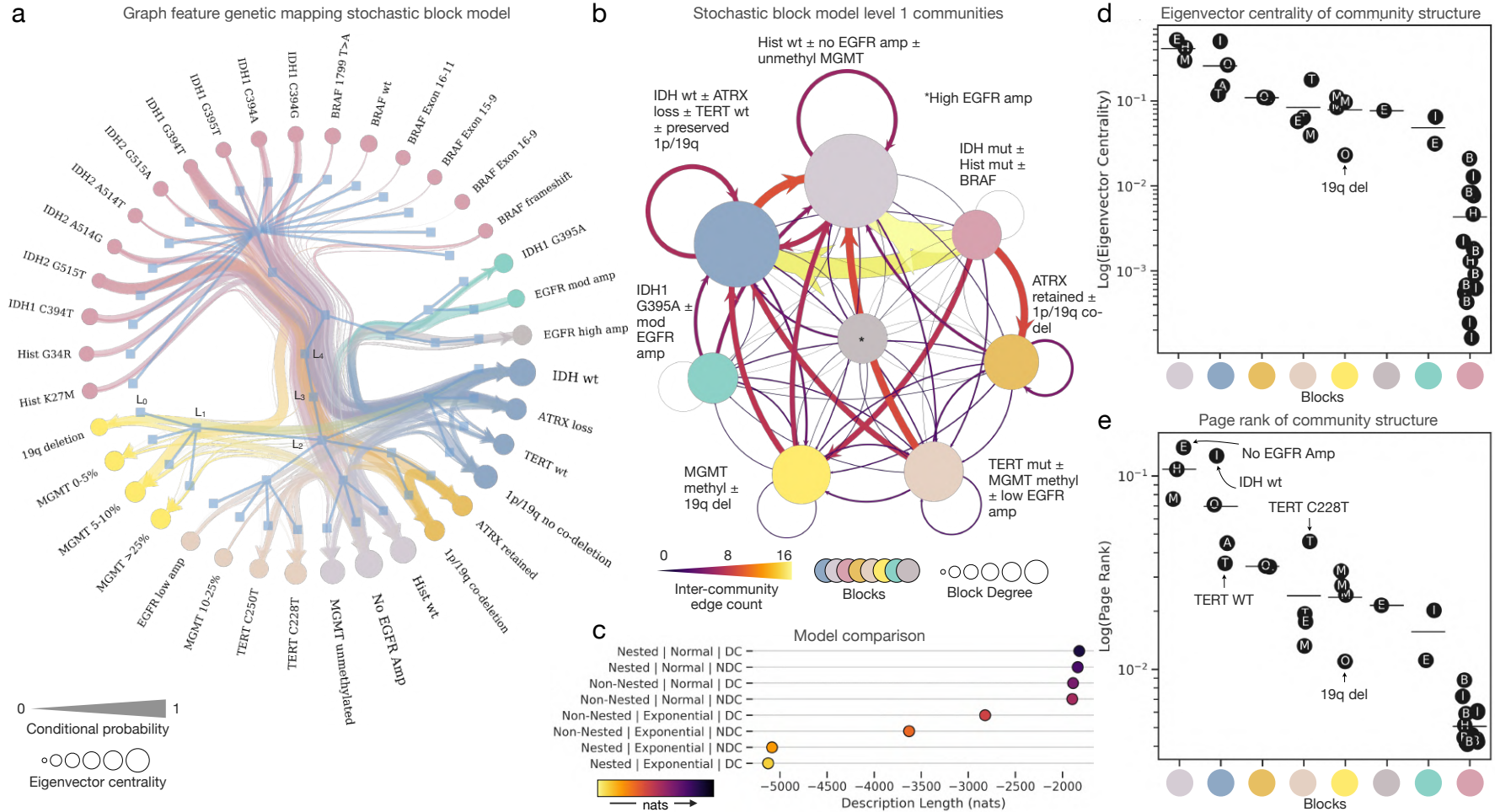

Supplementary Figure 4 – Graph feature genetic mapping identifies systematic brain tumour genetic links. a) Radial graph of layered, nested, degree-corrected, and exponentially weighted stochastic block model revealing the community structure of tumour genetics and their influence upon overall network topology. Communities are colour coded by the first level blocks of the hierarchical community structure. Edges are sized according to their conditional probability. Nodes are sized according to their weighted-eigenvector centrality. Hierarchical levels are annotated from level 0 ($L_0$) to level 4 ($L_4$). c) Fits in accordance with link probability by conditional probability, measured frequency and the comparative null

illustrate the description length of the layered model lower than the null, evidencing it a more suitable structure. c) Visualization of the first level hierarchy ($L_1$) with node colour as per that of panel a. Edge size and colour is proportional to the incidence of edges linking mutations between a given block. Node size is proportional to the degree of the corresponding block. d) There is a significant difference in weighted eigenvector centrality of tumour genetic factors when organized by stochastic block model community ($p<0.0001$). e) There is a significant difference in weighted page rank of tumour genetic factors when organized by stochastic block model community ($p<0.0001$). f) There is a significant difference in weighted hub centrality of tumour genetic factors when organized by stochastic block model community ($p<0.0001$). In panels d-e, block colour as per that of panel a, points are labelled by their corresponding abbreviation below and outliers within given communities are also annotated: A, ATRX; B, BRAF, E, EGFR; H, Histone; I, IDH; M, MGMT, O, 1p/19q; T, TERT.

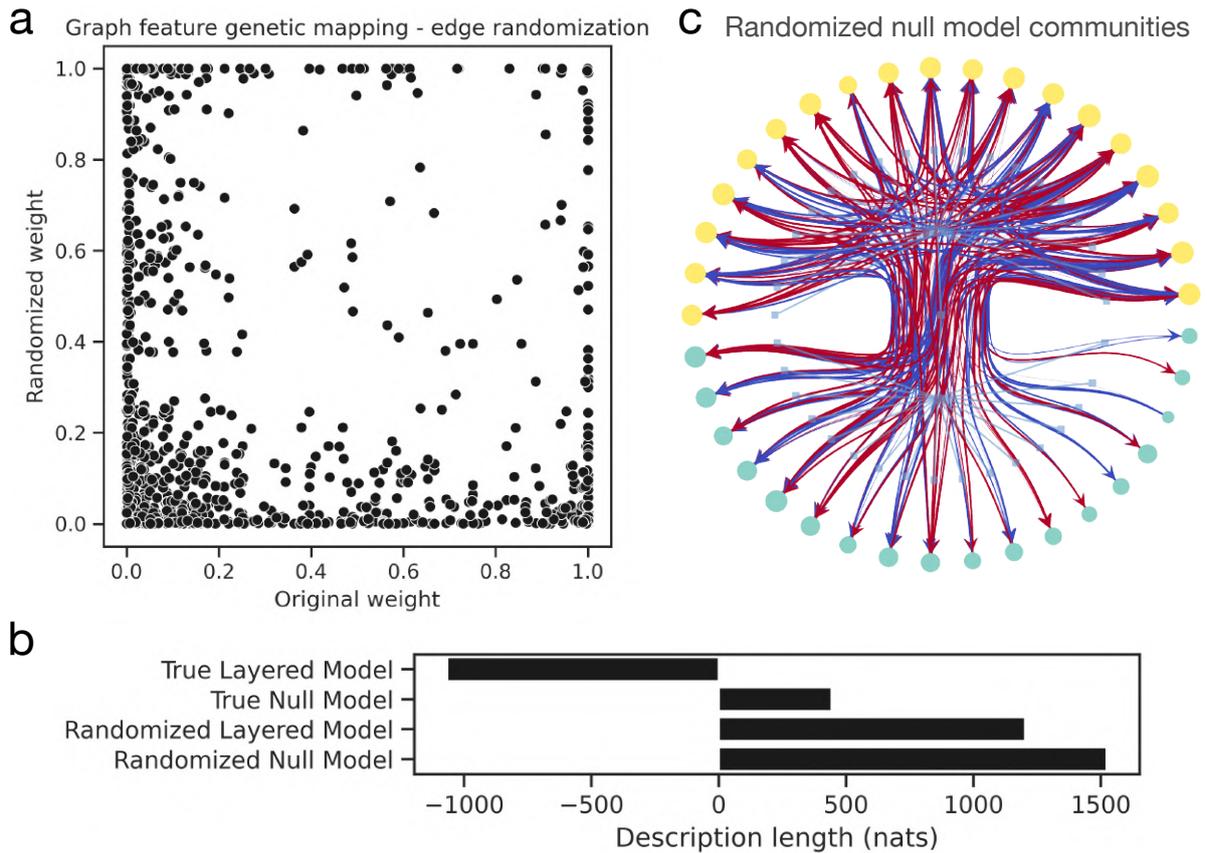

**Supplementary Figure 5 – Randomized null models of graph feature genetic mapping.** a) Scatterplot depicting edge randomization with original (true) edge weights along the x-axis, and those randomized along the y-axis. b) Both randomized layered and null-layered stochastic block models yield larger description lengths than true layered models, indicative of a poor fit. c) The community structure from edge-randomized models show no biologically meaningful (or plausible) segregation. Red edges depict randomized conditional probability edges, and blue edges from randomized sampling frequency. Since the randomized community structure is nonsensical, node labels have been withheld.

SEE HTML FILE

Download and open in web browser

**Supplementary Figure 6 - Interactive tumour genetic network.** Red edges indicate links between loci by conditional probability over and above sampling frequency effects. Blue edges indicate links between loci largely represented by sampling frequency. For conciseness, only the top 25% of edges are shown. The tab icon in top right of screen allows node labels, edge labels, edge, and node sizes to be modified based upon the variety of parameters fitted. Images can also be exported as static images.

SEE HTML FILE

Download and open in web browser

**Supplementary Figure 7 - Unthresholded interactive tumour genetic network.** Red edges indicate links between loci by conditional probability over and above sampling frequency effects. Blue edges indicate links between loci largely represented by sampling frequency. For a thresholded version of this network, we recommend review of static supplementary figure 2, or interactive supplementary figure 6. The tab icon in top right of screen allows node labels, edge labels, edge, and node sizes to be modified based upon the variety of parameters fitted. Images can also be exported as static images.

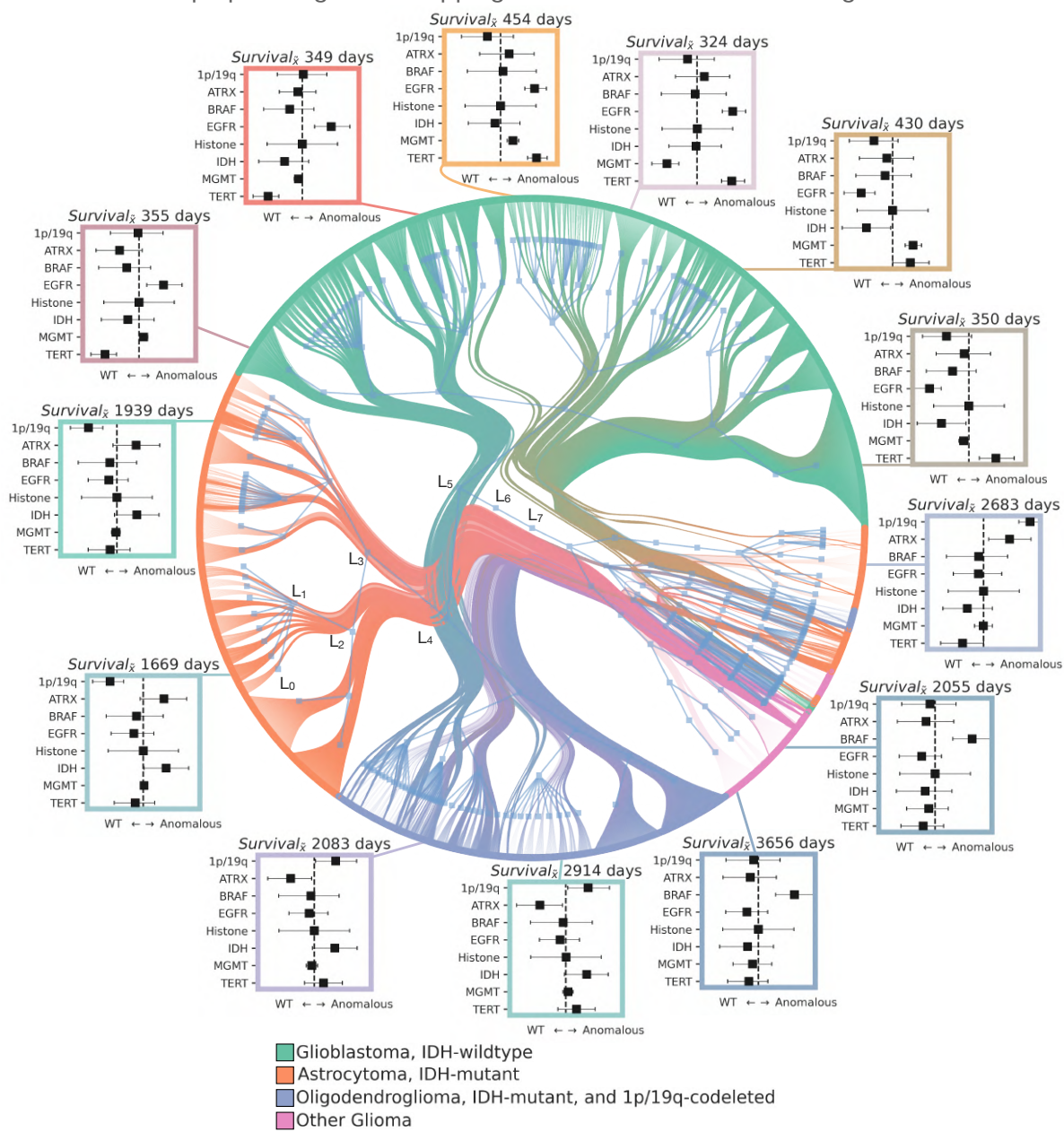

Supplementary Figure 8 – **Graph patient genetic mapping recovers overall diagnosis. Radial graph of nested, degree-corrected, and multivariate binomially weighted stochastic block model revealing the community structure of patients and the genetics of their brain tumour**. Patient nodes are color-coded by diagnosis, as per the colour-key. Around the radial graph are the breakdown of median survival and box and whisker plots for the coefficients and 95% confidence intervals of genetic loadings, where the coloured border of the plots depicts the corresponding community. All boxplots wherein the error-bar does not cross the vertical zero-line are significant, with features left of the vertical zero-line favouring the wild-type, and right of the zero-line favouring mutation. Note only the minimum spanning tree of the graph is shown, owing to visualisation constraints.

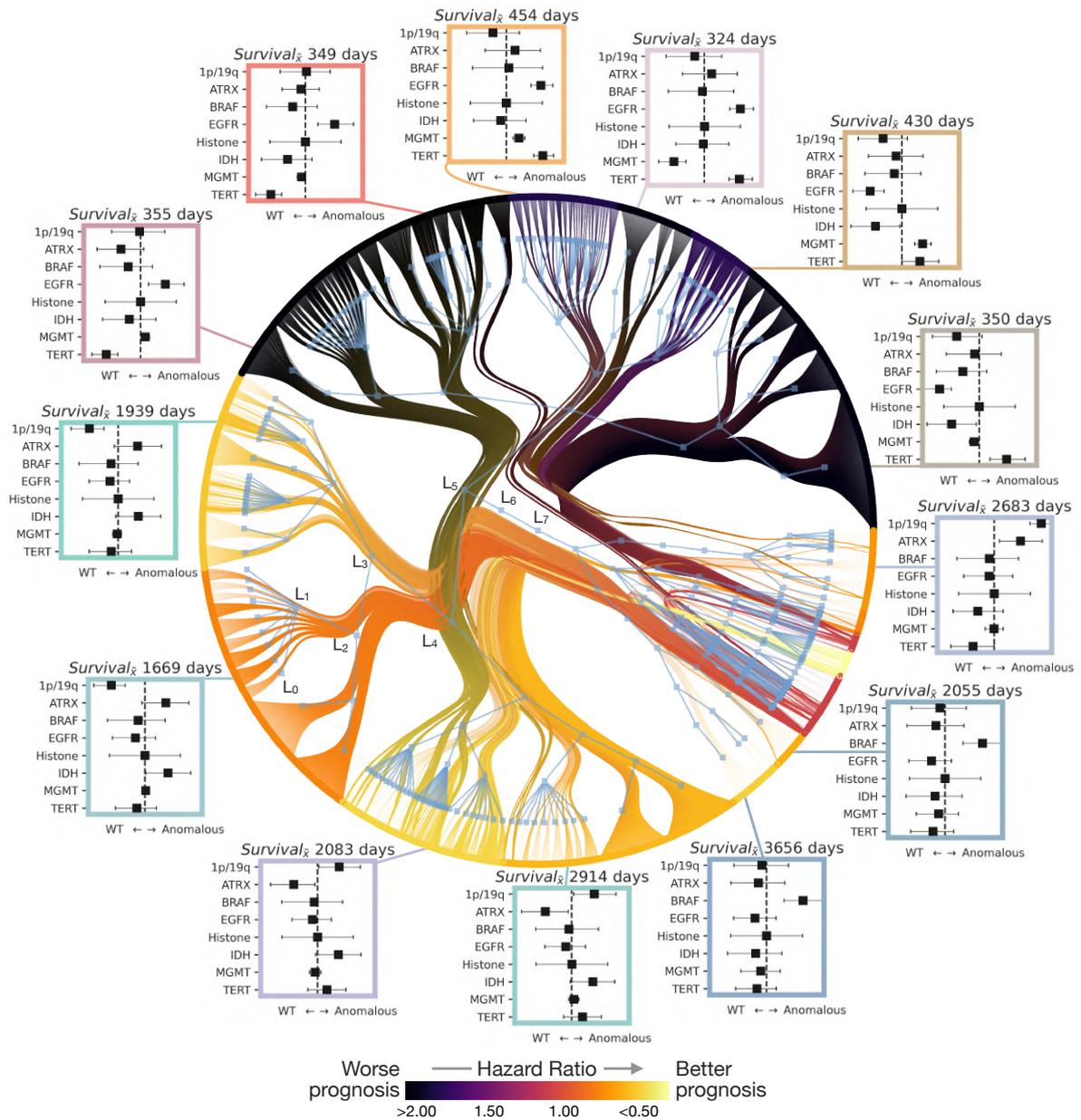

**Supplementary Figure 9 – Graph patient genetic mapping recovers offers more personalised survival predictions**. Radial graph of nested, degree-corrected, and multivariate binomially weighted stochastic block model revealing the community structure of patients and the genetics of their brain tumour. Patient nodes are color-coded by the hazard ratio of Cox's proportional hazard model. Around the radial graph are the breakdown of median survival and box and whisker plots for the coefficients and 95% confidence intervals of genetic loadings, where the coloured border of the plots depicts the corresponding community. All boxplots wherein the error-bar does not cross the vertical zero-line are significant, with features left of the vertical zero-line favouring the wild-type, and right of the zero-line favouring mutation. Note only the minimum spanning tree of the graph is shown, owing to visualisation constraints.

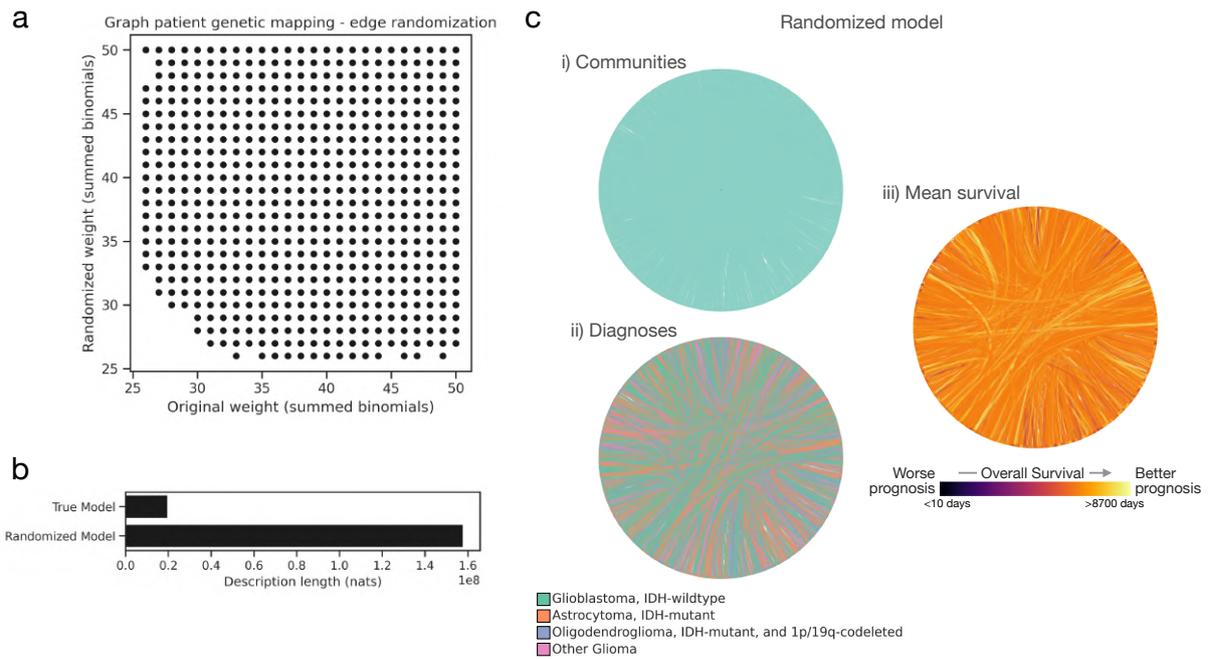

**Supplementary Figure 10 – Randomized null models of graph patient genetic mapping.** a) Scatterplot depicting edge randomisation with the sum of the original (true) edge weights along the x-axis, and the sum of those randomized along the y-axis. b) The randomized stochastic block model yields a far larger description lengths than the true model, indicative of a poor fit. c) There is no community structure from the randomized null model (panel i), bearing no relation to either diagnosis (panel ii), or mean survival (panel iii).

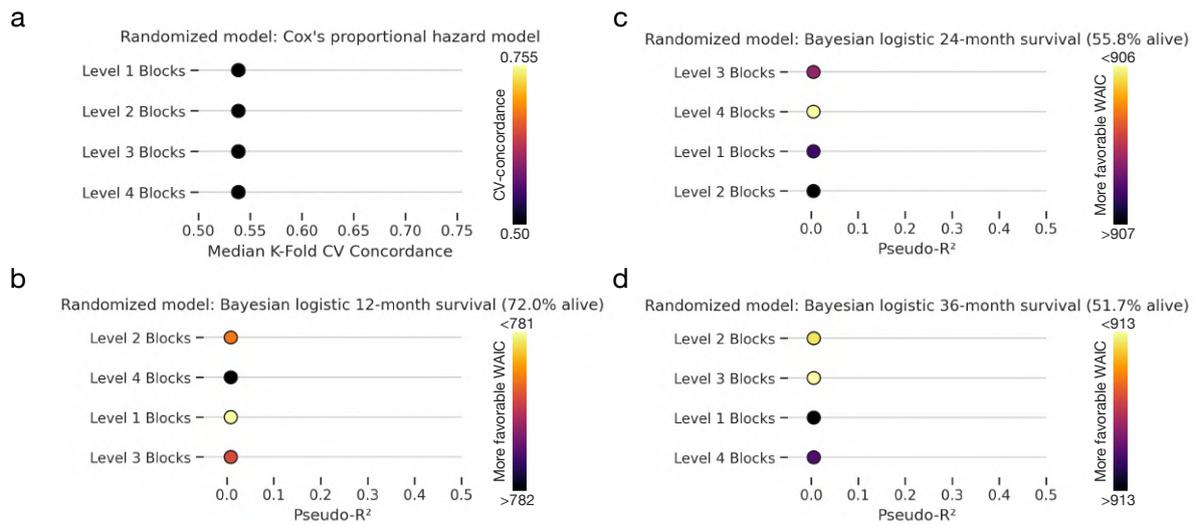

**Supplementary Figure 11 – Randomized null survival models.** a) Cox's proportional hazard, b) Bayesian logistic regression for 12-month, c) 24-month, and d) 36-month survival using the results of the randomized baseline model (Supplementary Figure 10) yields no predictive power in survival forecasting.